\documentstyle[12pt,english]{article}
\advance\topmargin by -1.6cm

\newenvironment{eq}
{\[\begin{array}}{\end{array}\]{}}

\let\rvec=\vec        

   \def\({\Bigl(}
\def\){\Bigr)}    \def\|{\Big|}
\def\then{~\Rightarrow~}   \def\o{\circ}
    \def\x{\times}
   \def\ox{\otimes}
 
\def\pl{{~\oplus~}}

\def\SUM{\displaystyle \sum}

\def\mid{\big\bracevert}

\def\subnoteq{\subset}

\def\supnoteq{\supset}
\def\and{\wedge}

\def\rin{{\,\in\kern-.42em\in}}
 
 \def\diag{{\,{\rm diag}\,}}
\def\sign{{\,{\rm sign}\,}}

\def\spec{\,{\rm spec}\,}
\def\rep{\,{\rm rep}\,}

\def\tr{{\,{\rm tr }\,}}
\def\det{\,{\rm det }\,}

\def\Int{\,{\rm Int}\,}

\def\ad{{\,{\rm ad}\,}}
\def\centr{\,{\rm centr}\,}
\def\Kern{\,{\rm kern}\,}

\def\sx{\rvec\x}
\def\weights{{{\bf weights}\,}}
\def\irrep{{{\bf irrep\,}}}
\def\rep{{{\bf rep\,}}}

\def\meas{{\bf meas}\,}

\def\A{{\,{\rm A\kern-.55emA}}}
\def\B{{\,{\rm I\kern-.2emB}}}
\def\C{{\,{\rm I\kern-.55emC}}}
\def\E{{\,{\rm I\kern-.2emE}}}
\def\G{{\,{\rm I\kern-.55emG}}}
\def\H{{\,{\rm I\kern-.2emH}}}
\def\I{{\,{\rm I\kern-.2emI}}}
\def\K{{\,{\rm I\kern-.2emK}}}
\def\L{{\,{\rm I\kern-.2emL}}}
\def\M{{\,{\rm I\kern-.16emM}}}
\def\N{{\,{\rm I\kern-.16emN}}}
\def\Q{{\,{\rm I\kern-.5emQ}}}
\def\R{{\,{\rm I\kern-.2emR}}}
\def\S{{\,{\rm I\kern-.42emS}}}
\def\T{{\,{\rm I\kern-.37emT}}}
\def\UU{{\,{\rm I\kern-.51emU}}}
\def\Z{{\,{\rm Z\kern-.32emZ}}}

\def\p{\partial}

\def\al{\alpha}  \def\be{\beta} \def\ga{\gamma}
\def\de{\delta}  \def\ep{\epsilon}  \def\ze{\zeta}
\def\th{\theta}   \def\vth{\vartheta} 
      \def\si{\sigma}
    
\def\phi{\varphi}  \def\Ga{\Gamma}  
    \def\La{\Lambda}

\def\vec#1{\underline{\bf vec}_{#1}}

\def\GL{{\bf GL}}  
\def\SL{{\bf SL}}
\def\U{{\bf U}} 
 
\def\O{{\bf O}}   
\def\SU{{\bf SU}} 
\def\SD{{\bf SD}} 
\def\SO{{\bf SO}}

 \def\D{{\bl D}}

\def\d#1{{\check{#1}}}
\def\angle#1{\langle#1\rangle}

\def\ro#1{{\rm #1}}
\def\bl#1{{\bf {#1}}}
\def\cl#1{{\cal #1}}
\def\ul#1{\underline{#1}}

\def\ol#1{\overline{#1}}

\def\com#1#2{\lbrack#1,#2\rbrack}

\def\acom#1#2{\{#1,#2\}}

\def\map{\longrightarrow}
\def\inmap{\hookrightarrow}
\def\lrmap{\leftrightarrow}

\def\mape{\longmapsto}

\begin{document}

\newpage\begin{titlepage} 
\hfill MPI-PhT/99-3


\vskip25mm
\centerline{\bf REPRESENTATIONS OF SPACETIME} 
\centerline{\bf AS UNITARY OPERATION CLASSES}
\vskip3mm
\centerline{\bf or}
\vskip3mm
\centerline{\bf AGAINST THE MONOCULTURE OF PARTICLE FIELDS}

\vskip1cm
\centerline{
Heinrich Saller\footnote{\scriptsize 
saller@mppmu.mpg.de}  
}
\centerline{Max-Planck-Institut f\"ur Physik and Astrophysik}
\centerline{Werner-Heisenberg-Institut f\"ur Physik}
\centerline{M\"unchen}
\vskip25mm

\centerline{\bf Abstract}

Spacetime is modelled as a homogeneous manifold given by the
classes of unitary $\U(2)$ operations in the general complex operations
$\GL(\C^2)$. The  residual representations of this 
noncompact symmetric space of rank two are characterized 
by two continuous real invariants,
one invariant interpreted as a  particle mass
 for a positive unitary subgroup and the second one for an indefinite 
unitary subgroup related to nonparticle interpretable
interaction ranges. Fields represent
 nonlinear spacetime $\GL(\C^2)/\U(2)$
by their quantization and include necessarily nonparticle contributions in 
the timelike part of their flat space Feynman propagator.


\end{titlepage}

{\small \tableofcontents}

\newpage

\section{Introduction}
\subsection{Some Historical Remarks}

Newton's interpretation of space and time having an
absolute ontology (two unaffected
 boxes wherein the physical objects play around) 
was by far more successful 
in the development of physical theories than Leibniz's opinion, who considered 
time and position space as relations, as labels to express their transformation
properties. With Einstein the two boxes became one spacetime box affected
by  and affecting the gravitational interaction.

Weyl\cite{WEYLRZM} made the first attempt to unify Einstein's gravity with Maxwell's
electrodynamics by 
explaining the electromagnetic interactions
as effected  by fields which connect and compatibilize 
spacetime dependent transformations from the  
noncompact abelian dilatation group $\D(1)=\exp \R$. 
This gauge idea, used for the wrong patient, 
was fruitful by switching over from
the noncompact $\D(1)$ to the apparently right patient, 
the compact abelian  transformation group $\U(1)=\exp
i\R$, a real Lie group defined in the complex. 
Therewith a dichotomy between external
spacetime  transformations with the Lorentz group $\O(1,3)$
and internal unitary transformations comprising
 the electromagnetic group $\U(1)$
 was  established. The
internal transformation group 
proliferated, the experimental and theoretical favourites being 
 today  the compact 
standard model interaction gauge groups
$\U(1)$, $\SU(2)$ and $\SU(3)$ for hypercharge, isospin and colour resp.

General relativity and electrodynamics came originally in
real formulations,
whereas quantum theory with its `probability amplitudes', 
characteristic phase relations (transition elements) 
and $\U(1)$-invariant scalar product was born as a complex 
theory. 
The gauge approach ties the electromagnetic
interaction to the  $\U(1)$-phases of
complex  matter fields.
The  complex representation of the internal 
compact real Lie group   
does not fit easily in a real representation structure of the external
transformations with the Poincar\'e group, i.e. the
vector spaces $\R^3$ and $\R^4$ 
for position space and spacetime translations resp. acted on by the 
 rotation and the Lorentz group, $\O(3)$ and $\O(1,3)$ resp.
But complex representations came rather early also
for the real spacetime transformations:
The twofold split of a ray with silver atoms in 
the original  Stern-Gerlach experiment
was the starting point to replace the rotation group $\SO(3)$ with their
real    irreducible representation spaces, necessarily odd dimensional, e.g. 
real 3-di\-men\-sio\-nal position space, by its 
twofold covering 
spin group $\SU(2)$. For the Lorentz group, this entailed the
transition to the complex represented real\footnote{\scriptsize 
The subindex $\R$ at the complex numbers $\C_\R=\R\pl i\R$ indicates its use 
as a complex
represented real structure. The complex 3-di\-men\-sio\-nal Lie group
$\SL(\C^2)$ and the real 6-di\-men\-sio\-nal one $\SL(\C_\R^2)$ are kept apart by
this - perhaps overcautious - notation.} Lie
group $\SL(\C_\R^2)$ covering the or\-tho\-chro\-nous group $\SO^+(1,3)$.
In a rather loose
external-internal `unification' both  groups, 
 the  Lorentz  $\SL(\C_\R^2)$ and the electromagnetic
 $\U(1)$ transformations come together 
 as subgroups of the full real 8-di\-men\-sio\-nal group $\GL(\C_\R^2)$
 with a central correlation \cite{S97}.  
This group is represented directly by
the transformations of charged spinor fields, e.g. of the right handed
lepton isosinglet fields in the standard model of electroweak and strong
interactions. In contrast to 
such an external-internal unification
of the Lorentz group with   the abelian hypercharge $\U(1)$
in $\GL(\C_\R^2)$ a unification of the nonabelian  groups $\SU(2)$ and $\SU(3)$
for isospin and colour with $\SL(\C^2_\R)$
remains an open problem\cite{S981,S982}.  

\subsection{Equations of Motion?}

The replacement of a finally oriented causality in Aristotelian physics by
time derivative equations of motions with initial conditions was,
as a method,  the most important progress initiated by Newton.
Subsequently, equations of motion were derived 
from Hamiltonians and Lagrangians using extremal principles.
In the course of the last century  Hamiltonians and Lagrangians were
more and more motivated 
and constructed as invariants with respect to transformation Lie groups 
and Lie algebras. 

In quantum mechanics 
the operational structures of physics come into full bloom\cite{FINK}:
The equations of motions can be
interpreted as the 
transformations with the Lie group $\exp t\in \D(1)$, modelling time,
expressed via the adjoint action 
${d\over dt}a=[iH,a]$ with a hermitian
Hamiltonian $H$ giving a  basis $iH$
for the time translation Lie algebra\footnote{\scriptsize
The Lie algebra for the Lie group $G$ is denoted as logarithm $\log G$.}  
$\log\D(1)=\R$. The 
action of time and its diagonalization is 
an algebraic eigenvalue problem $[H,a]= E(a) a$ - an equation of motion
is its differential formulation only, i.e. ${d\over dt}\cong i\ad H$.  

In the characteristic example of a  quantum harmonic  
oscillator the time action diagonalization gives  integer
energy  eigenvalues. The  
involved representation of time $\D(1)\map\U(1)$ 
by a unitary group establishes the probability structure since  $\U(1)$
is the invariance group of a scalar product for the complex
representation space.
Everything else, the definition of  
position and momentum
as real linear combinations of creation and annihilation operators,
which express the notion
'linear duality', the construction of a Hilbert space
with normalizable wave functions 
etc. are formulations for the basic $\U(1)$-re\-pre\-sen\-ta\-tion structure of time 
adapted for the  description of experiments in
the classical physics oriented language. 

The same procedure can be given explicitely e.g. for
the not so trivial  nonrelativistic hydrogen atom  
as done by Fock\cite{FOCK} using the rotation-perihel
invariance group of the Kepler
dynamics, i.e. $\SO(4)\cong{\SU(2)\x\SU(2)\over\{\pm 1\}}$ (compact, i.e.
definite unitary)
 for bound states and $\SO^+(1,3)\cong\SL(\C^2_\R)/\{\pm 1\}$ 
(real, but indefinite unitary)  
 for scattering states,
to determine the Hamiltonian as invariant and to give the
definite $\U(1)$ and indefinite 
$\U(1,1)$ unitarity structure resp. of the time action representations.

So far in quantum field theory, a replacement of the equations of motion,
e.g. in the standard model, by a purely algebraic 
transformation theory with eigenvalues and eigenvectors -
as seems appropriate for a quantum theory - which cannot only
describe the scattering of particles but also 
derive, in a bound state structure, 
 their existence and their  properties in terms of eigenvalues
has not succeeded yet.
In the following, I shall try some steps on this route.

\subsection{The Particle Prejudice}

Relativistic  quantum field theory is often praised as a progress  
in so far as  interactions and particles are unified - 
all interactions are parametrizable by particle fields.
Even if such a viewpoint is qualified by extending the particle language also 
to off-shell energy-momenta,
i.e. for mass $m$ particles to energy-momenta $q$ with
$q^2\ne m^2$, it is simply  not true. 
Apart from quarks and gluons as strong interaction parametrizing fields
postulated  without
particle asymptotics (confinement), the most prominent example is
the classical spinless Coulomb interaction
which comes in the quantum electromagnetic 
Lorentz vector field 
$\bl A(x)={\scriptsize\pmatrix{
\bl A_0+\bl A_3&\bl A_1-i\bl A_2\cr 
\bl A_1+i\bl A_2&\bl A_0-\bl A_3\cr}}(x)$ with four
components. The $\SO^+(1,3)$-Lorentz vector properties
with maximal abelian subgroup $\SO(2)\x\SO^+(1,1)$
leads to a unitary $\U(2)\x\U(1,1)$-'metric'.
As seen in the harmonic analysis with energy-momentum 
dependent creation and annihilation operators 
only the two transversal components, related to 
a $\U(2)$-scalar product  are particle
 interpretable as
left and right circularily polarized photons. From the remaining two
components with indefinite $\U(1,1)$-sesquilinear form  one  component
 is related to the gauge degree of freedom,
and the last 4th degree
of freedom describes a quantum field interaction without 
particle parametrization\cite{SBH95}.

An unreflected one to one correspondence of quantum fields with particles
is similar and 
somewhat  related to a superficial naive  interpretation of
Lorentz transformations for  spacetime translations 
$x={\scriptsize\pmatrix{
x_0+x_3&x_1-ix_2\cr x_1+ix_2&x_0-x_3\cr}}$
as blurring the difference
between time and position space. Obviously,
the situation is more subtle. Also 
in special relativity, timelike and spacelike
translations $\det x=x^2>0$ and $x^2<0$ resp.
are Lorentz operation compatible concepts - they are clearly distinct, 
but no longer linear
subspaces. 
The relativity of time translations $\R$ and position space translations
$\R^3$ can be seen in the homogeneous nonlinear structure of 
the absolute concepts `timelike' and `spacelike'. 
With the fixgroups (`little groups') 
$\SO(3)$, $\SO(1,2)$ and the semidirect $\SO(2)\sx\R^2$
for timelike, spacelike and lightlike translations resp. and
the dilatation group $\D(1)=\exp\R$ one has the nonlinear manifolds
for the nontrivial spacetime translations
\begin{eq}{lll}
\hbox{timelike future (past)}:&
\D(1)\x\SO^+(1,3)/\SO(3)&\cong\GL(\C^2_\R)/\U(2)\cr  
\hbox{spacelike}:&
\D(1)\x\SO^+(1,3)/\SO(1,2)&\cong\GL(\C^2_\R)/\U(1,1)\cr  
\hbox{lightlike future (past)}:&
\SO^+(1,3)/\SO(2)\sx\R^2&\cong\SL(\C^2_\R)/\U(1)\sx\C_\R\cr  
\end{eq}

The properties of free particle fields
are encoded in Feynman propagators\footnote{\scriptsize
The translation compatible shorthand (anti)commutator notation
$[a,b]_\pm(x-y)=[a(y),b(x)]_\pm$ is used.}, e.g. for
a hermitian scalar particle field $\bl\Phi$ with mass $m$
\begin{eq}{l}
\angle{\acom{\bl\Phi}{\bl\Phi}(x)-\ep(x_0)[\bl\Phi,\bl\Phi](x)}=
{i\over \pi}
\int {d^4q\over(2\pi)^3}{1\over q^2+io-m^2}e^{xiq}
\end{eq}with the on-shell quantization causally supported, i.e.
$[\bl\Phi,\bl\Phi](x)=0$ for $x^2<0$.
If one uses a rest system in linear spacetime and, 
therewith, a basis for  time translations in an  obviously not Lorentz 
compatible decomposition in time and position space,
timelike  translations $(x_0,\rvec x)$ with $x^2>0$ have in general 
also a nontrivial linear 
position space component $\rvec x$.
In relativistic field theories, 
the position space dependence of
nonrelativistic interactions like
the Yukawa or Coulomb interaction does not
arise from spacelike translations, but from timelike ones
$x^2>0$, in the example above from the off-shell causal contribution 
involving the principal value integration $\ro P$ 
\begin{eq}{c}
\ep(x_0)[\bl\Phi,\bl\Phi](x)=
\int {d^4q\over(2\pi)^3}\ep(x_0q_0)\de(m^2-q^2)e^{xiq}=
{1\over i\pi}
\int {d^4q\over(2\pi)^3}{1\over q_\ro P^2-m^2}e^{xiq}\cr
-2i\pi\int dx_0 \ep(x_0)[\bl\Phi,\bl\Phi](x)={e^{-|\rvec xm|}\over |\rvec x|}
\end{eq}Only the on-shell Fock value of
the quantization opposite commutator, in the example above 
\begin{eq}{l}
\angle{\acom{\bl\Phi}{\bl\Phi}(x)}
=\int {d^4q\over(2\pi)^3}\de(m^2-q^2)e^{xiq}
\end{eq}which is also spacelike supported, 
is relevant for the asymptotic particle interpretation.
The causally supported off-shell part
$\ep(x_0)[\bl\Phi,\bl\Phi](x)$ in the Feynman propagator is  
a particle related contribution to a more complicated 
spacetime representation structure as will be  elaborated in the 3rd chapter.
I think that relativistic quantum theory 
using particle related fields only is 
incomplete and unsatisfactory 
with respect to its causal spacetime representation content.

\subsection{The Complication of Spacetime Theories} 

One may ask for reasons why an algebraization of spacetime theories
is so difficult. One reason may be
that in the double dichotomy `abelian-nonabelian' and `compact-noncompact' 
\begin{eq}{c}
\begin{array}{|c||c|c|c|}\hline
&\hbox{abelian}&\hbox{nonabelian}&\hbox{eigenvalues, invariants}
\cr\hline\hline 
\hbox{compact}&\U(1)&\U(2)&\Q\cr\hline
\hbox{noncompact}&\GL(\C_\R)&\GL(\C^2_\R)&\R\cr\hline
\begin{array}{c}
\hbox{homogeneous}\cr
\hbox{(noncompact)}\end{array}
&\D(1)&\GL(\C^2_\R)/\U(2)&\R\cr\hline
\end{array}
\end{eq}seen in parallel to the physical concepts
\begin{eq}{c}\hskip-3mm
\begin{array}{|c||c|c|c|}\hline
&\hbox{abelian}&\hbox{nonabelian}&\hbox{quantum numbers}\cr\hline\hline 
\begin{array}{c}
\hbox{internal}\cr
\hbox{(compact)}\end{array}
&\hbox{electromagnetic}&\hbox{electroweak}
&\begin{array}{c}
\hbox{winding, charge numbers}\cr
\hbox{spin, multiplicities}\end{array}
\cr\hline
\begin{array}{c}
\hbox{external}\cr
\hbox{(noncompact)}\end{array}&
\hbox{time}&\hbox{spacetime}
&\begin{array}{c}
\hbox{frequencies, energies}\cr
\hbox{masses, interaction ranges}\end{array}
\cr\hline
\end{array}
\end{eq}spacetime operations are both nonabelian and noncompact.

The nondecomposable representations of compact-abelian transformations 
are
complex 1-di\-men\-sio\-nal, of compact-nonabelian transformations 
complex finite dimensional, 
both with rational eigenvalues\cite{LIE48,FULHAR}
 - as physical
properties called e.g. winding, charge or spin numbers.
As for the noncompact transformations,
the irreducible representations in the abelian case 
are still complex 1-di\-men\-sio\-nal,
in the nonabelian case 
in general infinite dimensional\cite{GELVIL,GELNAI}, 
in  both cases   with a continuous spectrum -
as physical properties called energies (frequencies) masses or interaction
ranges.

If we insist in the causality
compatible or\-tho\-chro\-nous Lorentz group $\SO^+(1,3)$ 
we have to face the representation complications of 
noncompact-nonabelian transformations.

\section{Spacetime as  Transformations}

In this chapter a  model for spacetime is proposed with
spacetime points as classes of transformations.

\subsection{A Mathematical Remark on `Naturalness'}

In mathematics, there exist `natural' structures 
connected with the solution of  `universal' problems\cite{ENS4} 
which may be superficially characterized as follows: 
A given structure gives rise to 
new ones by considering its internal  relations,
e.g. its selftransformations as binary relations.

Some well known elementary examples: 
Each abelian semigroup with cancellation rule
is naturally extendable to a unique group structure. This is 
used for the extension of
the natural numbers  to the integer ones 
as binary internal  relations 
modulo an addition $+$ induced equivalence
$\stackrel{+}{\sim}$ 
\begin{eq}{l}
\Z={\N\x\N\over \stackrel{+}{\sim}}
\hbox{ with }
(n_1,n_2)\stackrel{+}{\sim}(m_1,m_2)\iff n_1+m_2=m_1+n_2
\end{eq}or for the extension of the integers as ring to the rationals
as its unique  field structure with  a 
multiplication $\cdot$ induced equivalence
$\stackrel{\cdot}{\sim}$ 
\begin{eq}{l}
\Q={\Z\x[ \Z\setminus\{0\}]\over \stackrel{\cdot}{\sim}}
\hbox{ with }
(z_1,z_2)\stackrel{\cdot}{\sim}(u_1,u_2)\iff z_1u_2=u_1z_2
\end{eq}

By considering 
Cauchy series as countably infinite relations, each metrical space 
has its unique naturally Cauchy completed space. This is used for the
extension of the rationals $\Q$ 
with their natural order induced metric 
to the reals $\R=\Q^{\aleph_0}/\stackrel{C}{\sim}$
with a Cauchy series induced equivalence $\stackrel{C}{\sim}$. 

Another example is  the 
natural structure of multilinearity: Each vector space gives rise 
to a unique  unital associative algebra structure, its tensor algebra.
Different quotient algebras can be related to the algebras used in
classical and quantum theories\cite{S922}.

\subsection{Adjoint Transformation Structures}

Some natural concepts involving binary internal relations are called 
{\it adjoint}. They play a paramount role in physical theories,
not only for  gauge fields.
With respect to real and complex Lie transformation groups and 
algebras (always finite dimensional if not stated explicitly otherwise)
such adjoint concepts describe the action
of the transformations on themselves and lead
to characteristic doublings.

The adjoint doubling will be illustrated with the
example of the real 3-di\-men\-sio\-nal position space
whose translations, formalized by a
vector space $\R^3$, with the action of a rotation group $\SO(3)$
constitute a Euclidean semidirect product group
\begin{eq}{l}
\SO(3)\sx\R^3 \hbox{ with product }(O_1,\rvec x_1)\o (O_2,\rvec x_2)
=(O_1O_2,\rvec x_1+O_2(\rvec x_2))
\end{eq} 

$\SO(3)\sx\R^3$ is an example for an {\it adjoint affine Lie group}
where, in general, a Lie group $G$ is represented in the automorphisms of the vector space
structure of its Lie algebra $\log G$
\begin{eq}{l}
G\x\log G\map\log G,~~(g,x)\mape \Int g(x)=g\o x\o g^{-1}\cr
\Int g_1\o \Int g_2=\Int g_1g_2
\end{eq}The adjoint group
representation is faithful for the {\it adjoint group} $\Int G$,
defined by the classes of the group elements with respect to the centrum,
i.e. the kernel of the group representation $\Int$
\begin{eq}{l}
\Int G\sx\ul{\log G},~~
\Int G= G/\centr G\cr
\hbox{with product }(g_1,x_1)\o(g_2,x_2)=(g_1g_2,x_1+\Int g_1(x_2))\cr
\end{eq}The `linear underlining' of the Lie algebra $\ul{\log G}$
indicates that only its linear vector space structure is relevant
for this adjoint doubling, the Lie bracket
of the 2nd factor has to be `forgotten'.

The Euclidean group for position space, mentioned above,
is the adjoint affine group
of the spin group $\SU(2)$
\begin{eq}{l}
\SO(3)\sx\R^3=\Int\SU(2)\sx\ul{\log\SU(2)},~~\centr\SU(2)=\{\pm\bl 1_2\}\cr
\hbox{with product }(u_1,x_1)\o(u_2,x_2)=(u_1u_2,x_1+u\o x_2\o u^*)
\end{eq}

Starting from the defining and fundamental complex 2-di\-men\-sio\-nal Pauli
$\SU(2)$-re\-pre\-sen\-ta\-tion  by $u=\exp i\rvec\al\rvec\si$ with spin
$J={1\over2}$
which - up to equivalence - gives all irreducible $\SU(2)$-re\-pre\-sen\-ta\-tions
$[2J]$, $J=0,{1\over2},1,\dots$ 
with dimension $(1+2J)$ by totally symmetrical tensor products,
one obtains the position space translations with the adjoint spin 
representation $[2]$
as vector space structure of the spin Lie algebra,  
in a Leibnizian interpretation as binary 
relations (traceless linear mappings) for Pauli spinors
\begin{eq}{l}
\hbox{position space translations}=\ul{\log\SU(2)}\cong\R^3\cr
\ul{\log\SU(2)}=\{x:\C_\R^2\map\C_\R^2\mid 
\tr x=0,~x= x^*=\rvec\si\rvec x={\scriptsize\pmatrix{
x_3&x_1-ix_2\cr x_1+ix_2&-x_3\cr}}\}
\end{eq}The rotations are realized by the adjoint representation
\begin{eq}{l}
u\in\SU(2):~~ x\mape u\o x\o u^*= O(u)(x)
\end{eq}The position space metric 
comes as negative definite Killing form, 
inherited from the spin Lie algebra,
i.e.
the $\SU(2)$-invariant double trace  $\tr x\o y$,
 with the quadratic form
as the determinant $x^2={1\over2}\tr x\o  x=-\det x$.

For a Lie algebra $L$, the {\it adjoint affine Lie algebra} is defined
by the adjoint representation
 which realizes the Lie bracket by the commutator
of the endomorphisms of its vector space structure
\begin{eq}{l}
L\x L\map L,~~(l,x)\mape \ad l(x)=\com lx\cr
\ad\com{l_1}{l_2}=\com{\ad l_1}{\ad l_2}
\end{eq}Only the adjoint Lie algebra  
$\ad L$ given by the classes of the Lie algebra with respect to the centrum 
is faithfully represented.
The adjoint affine Lie algebra is as vector space the 
direct sum $\ad L\pl L$ and as Lie algebra 
the semidirect bracket product, denoted by the direct sum-semidirect Lie
bracket symbol $\rvec\pl$ 
\begin{eq}{l}
\ad L~\rvec\pl~\ul L=\{l+x\mid l,x\in L\},~~\ad L= L/\centr L
\cr
\hbox{with bracket }[l_1+x_1,l_2+x_2]=[l_1,l_2]+\ad l_1(x_2)-\ad l_2(x_1)
\end{eq}The second factor $\ul L$ 
in this adjoint doubling is the vector space structure
of the Lie algebra.

For Euclidean position space,
the adjoint affine Lie algebra for the angular momenta
$\log\SO(3)$ is the Lie algebra of the Euclidean group
\begin{eq}{l}
\log\SO(3)\rvec\pl\R^3
\cong\log\SU(2)\rvec\pl\ul{\log\SU(2)}
\end{eq}

Both adjoint doublings,  
the adjoint affine group and the adjoint affine Lie algebra, are related 
to the realization
of a group $G$ on itself by inner automorphisms
\begin{eq}{l}
 G\x G\map G,~~(g,a)\mape \Int g(a)=gag^{-1}\cr
\Int g_1\o \Int g_2=\Int g_1g_2
\end{eq}leading to the {\it adjoint group doubling} 
as the semidirect product
\begin{eq}{l}
\Int G\sx G=\{(g,a)\mid g,a\in G\}\cr
\hbox{with product }
(g_1,a_1)\o(g_2,a_2)=(g_1g_2,a_1\Int g(a_2)) 
\end{eq}Each semidirect group $G'\sx G$ is isomorphic to a subgroup of
the adjoint group doubling $\Int G\sx G$ which is universal in this sense.  

The adjoint doubling of the spin group is its semidirect product
with the rotation group
\begin{eq}{l}
\SO(3)\sx\SU(2)
\end{eq}

\subsection{Spacetime and the Causal Poincar\'e Group}

For a Lie group $G$ with Lie algebra $\log G$ one has the three
semidirect adjoint doublings, reflecting two steps of
infinitesimalization
\begin{eq}{c}
\begin{array}{|c|c|c|}\hline
&\hbox{name}&\hbox{example}\cr\hline\hline
\Int G\sx G&\hbox{adjoint group doubling}&\SO(3)\sx\SU(2)\cr\hline
\Int G\sx\ul{\log G}&\hbox{adjoint affine group}&\SO(3)\sx\R^3\cr\hline
\log\Int G\rvec\pl\ul{\log G}&\hbox{adjoint affine Lie algebra}
&\log \SO(3)\rvec\pl\R^3\cr\hline\end{array}
\end{eq}They were discussed in the last section for the unitary spin group
$u^*=u^{-1}\in\SU(2)$ with the rotations $\SO(3)$ 
as adjoint group and
the position space translations $\R^3$ as vector space
structure of the spin Lie algebra $\log\SU(2)$.

What about spacetime?
The spacetime  translations $\R^4$
(Minkowski space) with the or\-tho\-chro\-nous Lorentz group
action constitute the semidirect product  Poincar\'e group
\begin{eq}{l}
\SO^+(1,3)\sx\R^4
\end{eq}The Poincar\'e group is not the 
adjoint affine Lie group of the real 6-di\-men\-sio\-nal
Lie group $\SL(\C_\R^2)$
\begin{eq}{rl}
\Int\SL(\C_\R^2)\sx\ul{\log \SL(\C_\R^2)}&=\SO^+(1,3)\sx\R^6\cr
\Int \SL(\C_\R^2)&=\SL(\C_\R^2)/\{\pm\bl1_2\}\cong\SO^+(1,3)
\end{eq}This
real 12-di\-men\-sio\-nal 
group is relevant for the gauge structures in Min\-kow\-ski space
 where the curvature fields, e.g. the electromagnetic field strenghts
 $\{F^{jk}=-F^{kj}\}_{j,k=0}^3=\{\rvec E,\rvec B\}$,
represent the real 6-di\-men\-sio\-nal vector space structure of
the Lorentz Lie algebra with the adjoint Lorentz group action.

At first sight it seems unnatural to relate 
the real 4-di\-men\-sio\-nal Minkowski translations $\R^4$ to the 
real 6-di\-men\-sio\-nal Lorentz Lie algebra $\log\SO^+(1,3)\cong\log\SL(\C_\R^2)$.
However, it is exactly the complex 
representation  of the real covering group $\SL(\C_\R^2)$
which makes this relation natural in the mathematical sense.
 Only in this context, the Poincar\'e group 
for flat spacetime can be understood  
as arising from an adjoint doubling, i.e. related to internal
 relations of a transformation group.  

In the case of 
complex represented real transformations there are two kinds of adjoint structures.
It may be helpful to give the construction first in abstract terms:
If a semigroup $G$ has a reflection (conjugation), i.e. an involutive 
contra-automorphism defined by
\begin{eq}{l}
*:G\map G,~~g^{**}=g,~~(gh)^*=h^*g^*
\end{eq}it defines its {\it $*$-symmetric domain} as the subset
\begin{eq}{l}
D(G)=\{d\in G\mid d^*=d\}
\end{eq}The concatenation of the inversion of a group $G$ as canonical
group reflection with any  reflection (conjugation) $*$ is an
involutive automorphism
\begin{eq}{l}
\hat{~}:G\map G,~~\hat g=(g^{-1})^*=(g^*)^{-1}
\end{eq}The invariants for this automorphism constitute the
$*$-unitary subgroup
\begin{eq}{l}
U(G)=\{u\in G\mid u^{-1}=u^*\}
\end{eq}For a group with conjugation both the symmetric domain $D(G)$
and the unitary subgroup $U(G)$ can be used
for adjoint structures.

Physically relevant examples used in the following  are the full general complex
linear groups
$\GL(\C_\R^n)$,
considered as real Lie groups and
definable by  the nonsingular complex $n\x n$-matrices
with the hermitian matrix conjugation $*$.
They will be used for time in the case $n=1$ and for spacetime  
with $n=2$. The group $\GL(\C^n_\R)$ has 
the real $n^2$-di\-men\-sio\-nal 
 submanifold $\D(n)$ as its symmetric domain and 
the real $n^2$-di\-men\-sio\-nal Lie subgroup $\U(n)$ as 
the group with the invariants
\begin{eq}{ll}
\D(n)&=\{d\in\GL(\C_\R^n)\mid d^*=d\},~~
\U(n)=\{u\in\GL(\C_\R^n)\mid u^*=u^{-1}\}\cr
\end{eq}The symmetric domain is a symmetric space\cite{HEL}
with the maximal compact group as fixgroup. It is
the direct product of the abelian group $\D(\bl1_n)=\bl1_n\exp\R$
and the globally symmetric space $\SD(n)=\SL(\C_\R^n)/\SU(n)$
\begin{eq}{l}
\D(n)\cong\GL(\C_\R^n)/\U(n)\cong\D(\bl1_n)\x
\SL(\C_\R^n)/\SU(n)\cr
\end{eq}

Back to the general structure:
A group $G$ with  two reflections,
$g\lrmap g^*$ (conjugation) and  $g\lrmap g^{-1}$ (inversion), gives
rise to two types of
adjoint doublings.
The inversion induced inner automorphisms of the group $G$ described
in the former section 
\begin{eq}{l}
 G\x G\map G,~~(g,a)\mape 
  \Int g(a)=gag^{-1}=(\hat g a^*\hat g^*)^*\cr
\Int g_1\o \Int g_2=\Int g_1g_2,~~\Kern \Int G =\centr G 
\end{eq}are, in general, not compatible with the conjugation.
In addition and analogy to the inner automorphisms Int, the  group $G$
 allows
the conjugation compatible bijections, denoted by  $\Int_*$
\begin{eq}{l}
  G\x G\map G,~~(g,a)\mape \Int_* g(a)=gag^*=(ga^*g^*)^*\cr
\Int_* g_1\o \Int_* g_2=\Int_* g_1g_2
\end{eq}Also these  bijections constitute a realization of the group $G$
with the kernel defining the faithfully realized classes
$\Int_* G$
\begin{eq}{l}
\Kern\Int_*=\{h\in G\mid hgh^*=h\hbox{ for all }g\in G\}\cr
\Int_*G=G/\Kern\Int_*
\end{eq}For the unitary elements $u\in U(G)$ the
bijections coincide with the inner automorphisms,
i.e. $\Int_*u=\Int u$,  not, however,
in general. The analogue structure to 
the adjoint group doubling $\Int G\sx G$ 
 is given by 
the action of the conjugation compatible bijections on
the symmetric domain $D(G)$, which will be called the 
{\it adjoint symmetric transformation space}
\begin{eq}{l}
\Int_* G\sx_* D(G)
\end{eq}which, in general in contrast to $\Int G\sx G$, is no semidirect 
product group. 

The two types of adjoint doublings are illustrated for
the physically relevant examples $\GL(\C^n)$: One obtains 
for the complex case $\C$ 
with  the inversion and for the complex represented real one $\C_\R$
with the conjugation
\begin{eq}{llll}
\hbox{inversion:}&\Int\GL(\C^n)&=\GL(\C^n)/\GL(\C)&=\SL(\C^n)/\I(n)\cr
\hbox{conjugation:}&\Int_*\GL(\C_\R^n)&=\GL(\C_\R^n)/\U(\bl1_n)&=
\D(\bl1_n)\x\SL(\C_\R^n)/\I(n)\cr
\end{eq}with the cyclotomic group 
$\I(n)=\{z\in\C\mid z^n=1\}$ as $\SL(\C^n)$-centrum.
This leads to the adjoint group doublings and the 
adjoint symmetric transformation spaces
\begin{eq}{ll}
n=1:\hskip-3mm&\left\{\begin{array}{llll}
\hbox{inversion:}&\Int \GL(\C)\sx\GL(\C)&=\GL(\C)\cr
\hbox{conjugation:}&\Int_* \GL(\C_\R)\sx_*\D(1)&=\D(1)\sx_*
\D(1)\end{array}\right.\cr
n=2:\hskip-3mm&\left\{\begin{array}{llll}
\hbox{inversion:}&\Int \GL(\C^2)\sx\GL(\C^2)&=\SL(\C^2)/\I(2)\sx\GL(\C^2)\cr
\hbox{conjugation:}&\Int_* \GL(\C^2_\R)\sx_*\D(2)&=
[\D(\bl 1_2)\x\SO^+(1,3)]\sx_*\D(2)\end{array}\right.\cr
\end{eq}For spacetime with $n=2$ the conjugate
adjoint action involves the direct product of 
the or\-tho\-chro\-nous Lorentz group
and the dilatation group $\D(\bl1_2)$,
called causal group in this context.

Obviously for Lie symmetries the adjoint Lie group 
structures are linearizable
with Lie algebra structures, first in general:
For a Lie group $G$ with reflection (conjugation), 
the Lie algebra $\log G$ inherits
the reflection (conjugation). Therefore, it is the
direct sum of the  $*$-antisymmetrical
Lie subalgebra $l^*=-l$ as Lie algebra of the unitary Lie subgroup and the
isomorphic $*$-symmetrical vector subspace 
$x=+x^*$ as tangent structure of the symmetric manifold $D(G)= G/U(G)$
\begin{eq}{l}
\log G=\log G_-\pl\log G_+,~~\left\{\begin{array}{ll}
\log G_-&=\log U(G)\cr
\log G_+&\cong\log G/\log U(G)\end{array}\right.
\end{eq}In the example above one has
in addition to the Lie algebra $\log\U(n)$ 
as $\U(n)$-tangent space a real $n^2$-di\-men\-sio\-nal
vector subspace $\R(n)$ as  tangent space
of the symmetrical domain $\D(n)$ which, for $n=2$,  
will be used as spacetime translations (Minkowski space)
\begin{eq}{rl}
\log\GL(\C_\R^n)&=\log\U(n)\pl\R(n)\cr
\R(n)&\cong\log\GL(\C_\R^n)/\log \U(n)\cong\R^{n^2}
\end{eq}

In addition to the adjoint affine Lie group $\Int G\sx\ul{\log G}$
involving the inversion as  natural reflection
one has now also the conjugate adjoint representation of the group
on its Lie algebra 
 \begin{eq}{l} 
G\x\log G\map\log G,~(g,m)\mape\Int_*g(m)=g\o m\o g^*
=(g\o m^*\o g^*)^*\cr
\Int_* g_1\o \Int_* g_2=\Int_* g_1g_2
\end{eq}which, with the conjugation compatibility,  can be restricted to the 
symmetrical and antisymmetrical vector subspaces
of $\log G$. The
{\it conjugate adjoint affine Lie group}
is defined with the symmetrical subspace as translations 
\begin{eq}{l}
\Int_* G\sx_*\log G_+ 
 \end{eq}

With respect to the two adjoint doublings 
$\Int G\sx\ul{\log G}$ with inversion and 
$\Int_* G\sx_*\log G_+$ with conjugation 
one obtains in the spacetime relevant example ($n=2$) 
for the 2nd case the Poincar\'e group
with an additional causal group action 
\begin{eq}{ll}
n=1:\hskip-3mm&\left\{\begin{array}{llll}
\hbox{inversion:}&\Int \GL(\C)\sx\ul{\log\GL(\C)}&=\C\cr
\hbox{conjugation:}&\Int_* \GL(\C_\R)\sx_*\R(1)&=\D(1)\sx_*
\R\end{array}\right.\cr
n=2:\hskip-3mm&\left\{\begin{array}{llll}
\hbox{inversion:}&\Int \GL(\C^2)\sx\ul{\log \GL(\C^2)}
&\hskip-3mm=\SL(\C^2)/\I(2)\sx \C^4\cr
\hbox{conjugation:}&\Int_* \GL(\C^2_\R)\sx_*\R(2)&\hskip-3mm=
[\D(\bl1_2)\x\SO^+(1,3)]\sx_*\R^4\end{array}\right.\cr
\end{eq}

All finite dimensional  irreducible complex $\SL(\C_\R^2)$-re\-pre\-sen\-ta\-tions 
$[2L|2R]$ with halfintegers $L,R=0,{1\over2},1,\dots$
and dimension $(1+2L)(1+2R)$ 
can be built - up to equivalence - by the totally symmetrical tensor products
of the two fundamental Weyl representations, related to each other by the
conjugation induced automorphism
\begin{eq}{llll}
\hbox{left handed}&[1|0]\hbox{ by}&s&=\exp(+\rvec\be+i\rvec\al)\rvec\si\cr 
\hbox{right handed}&[0|1]\hbox{ by}
&\hat s&=\exp(-\rvec\be+i\rvec\al)\rvec\si\cr 
\end{eq}The representations have the conjugation 
$[2L|2R]^*=[2R|2L]$. The hermitian
irreducible $(1+2J)^2$-di\-men\-sio\-nal representations 
$[2J|2J]$ with $J=0,{1\over2},1,\dots$ are generated by the 
complex 4-di\-men\-sio\-nal Min\-kow\-ski $\SL(\C_\R^2)$-re\-pre\-sen\-ta\-tion
$[1|1]=[1|0]\ox[0|1]$ with 
the linear binary relations for Weyl spinors.
The 
symmetric (real) subspace
is the Cartan representation of the spacetime translations 
by linear spinor mappings with the Weyl matrices $\si^k\cong(\bl1_2,\rvec\si)$
\begin{eq}{l}
\hbox{spacetime translations}=
\log\GL(\C_\R^2)_+=\R(2)\cong\log\GL(\C_\R^2)/\log \U(2)\cr
\R(2)=\{x:\C_\R^2\map\C_\R^2\mid 
x=x^*=\si^kx_k=
{\scriptsize\pmatrix{
x_0+x_3&x_1-ix_2\cr x_1+ix_2&x_0-x_3\cr}}\}\cong\R^4
\end{eq}They are acted on with   
the Lorentz and the 
causal group  by the conjugate adjoint representation
\begin{eq}{ll}
s\in\SL(\C_\R^2):& x\mape s\o x\o s^*= \La(s)(x)\cr
d=d^*\in\D(\bl1_2):&x\mape d\o x\o d^*= D(d)(x)
\end{eq}The Lorentz metric
comes as product $g=\ep\ox\ep^{-1}$ with the
invariant spinor metric, i.e. the antisymmetric 
bilinear $\C^2$-volume form $\ep=-\ep^T$, leading to 
the indefinite signature $\sign g=(1,3)$.

\subsection{Spacetime as Unitary Operation Classes}

To summarize the salient structures of the last section
which will  be used in the following:
The conjugate adjoint operation structure
for the group $\GL(\C^n_\R)$   suggests the 
definition of nonlinear models for time and
spacetime as symmetric domains for complex linear transformations
where the spacetime points are the complex linear operations
modulo the maximal compact unitary operation group
\begin{eq}{rll}
&\D(n)&=\GL(\C^n_\R)/\U(n)\cr
\hbox{time ($n=1$):}&\D(1)&=\exp\R\cr 
\hbox{spacetime ($n=2$):}&\D(2)&\cong
\D(\bl1_2)\x\SO^+(1,3)/\SO(3) 
\end{eq}Time comes as group, spacetime as homogeneous manifold.
$\D(n)$ is the orientation manifold\cite{WEYLRZM} 
of scalar products in $n$ dimensions
\cite{S97}.

The translations are the corresponding tangent structures 
\begin{eq}{rll}
&\R(n)&=\log\GL(\C^n_\R)/\log\U(n)\cong\R^{n^2}\cr
\hbox{time translations ($n=1$):}&\R(1)&=\R\cr 
\hbox{spacetime translations ($n=2$):}&\R(2)&\cong \R\pl\log\SO^+(1,3)/\log\SO(3)
\end{eq}

As subsets of the complex $(n\x n)$-matrices which constitute a stellar algebra
time and spacetime carry the  spectrum induced order, i.e.
the natural order for time and  
the Min\-kow\-ski partial order for  spacetime
\begin{eq}{rl}
x=x^*=
{\scriptsize\pmatrix{
x_0+x_3&x_1-ix_2\cr x_1+ix_2&x_0-x_3\cr}}
\hbox{ positive}&\iff\spec x\ge0\cr
&\iff x^2=\det x\ge0,~~x_0={1\over2}\tr x\ge0\cr
&\iff x=\vth(x^2)\ep(x_0)x
\end{eq}

The conjugate  adjoint affine group is the semidirect 
 causal Poincar\'e group
\begin{eq}{l}
[\D(\bl1_2)\x\SO^+(1,3)]\sx_*\R(2)
\end{eq}Here in the conjugate adjoint doubling, the causal structure and the boost structure
arises twice - globally 
as $\D(\bl1_2)$ and $\SO^+(1,3)/\SO(3)$
and in the tangent space $\R(2)\cong\R\pl\R^3$
as time and position space translations where the
decomposition is incompatible with the
$\SO^+(1,3)$-action.

\section{Representations of Spacetime}

In analogy to Lie groups and algebras also  spacetime in the
symmetric space model $\D(2)=\GL(\C^2_\R)/\U(2)$ has linear representations.
These representations will be  constructed 
as residues in analogy to
the representations of time, modelled by the group $\D(1)=\exp\R$,
which is used in the quantization of the quantum mechanical basic dual 
pair 'position-momentum'.

\subsection{Quantum Representations of Time}

A dynamics is a representation of time, expressed in quantum mechanics
by the noncommutativity of the generating operators. 
In the simplest cases of a harmonic oscillator or of a free mass point
one has the time dependent commutation relations of the 
dual position-momentum pair $(\bl x,\bl p)$ which generates the
operator algebra
\begin{eq}{l}
{\scriptsize\pmatrix{
\com{i\bl p}{\bl x}&\com {\bl x}{\bl x}\cr
\com {\bl p}{\bl p}&\com {\bl x}{-i\bl p}\cr}}(t)=\left\{
\begin{array}{ll}
D({t\over M}| m ^2)= {\scriptsize\pmatrix{
\cos t m &{i\over M m }\sin t m \cr 
iM m  \sin t m &\cos t m \cr}}&
\hskip-3mm\begin{array}{l}\hbox{oscillator  mass $M$}\cr
\hbox{and frequency $ m $}\end{array}\cr
D({t\over M}|0)=
{\scriptsize\pmatrix{
1&{it\over M}\cr
0&1\cr}}
&\hskip-2mm\hbox{free point  mass $M$}\end{array}\right.
\end{eq}with the shorthand notation
$\com{a(s)}{b(t)}=\com ab(t-s)$, valid for all matrix elements.

The time translations
which generate the $\D(1)$-re\-pre\-sen\-ta\-tion
 are 
quantum represented with the Hamiltonian, 
e.g. for the  harmonic oscillator 
with creation and annihilation operator $(\ro u,\ro u^*)$
\begin{eq}{l}
H={\bl p^2\over2M}+{ m ^2M\over2}\bl x^2= m 
{\acom{\ro  u}{\ro u^*}\over2},~~\ro u={M m \bl x-i\bl p\over\sqrt{2M m }}\cr
\D(1)\ni e ^t\mape \com{\ro u^*}{\ro u}(t)= e ^{ti m} 
\in\U(1)\end{eq}The 
harmonic oscillator $\D(1)$-re\-pre\-sen\-ta\-tion by  
position-momentum is
 decomposable in two irreducible representations
 in $\U(1)\ni e^{\pm tim}$, dual to each other with the
 $\SO(2)$-metric
${\scriptsize\pmatrix{{1\over Mm}&0\cr 0 &Mm\cr}}$
built with the intrinsic oscillator length $\ell^2={1\over Mm}$
\begin{eq}{l} 
\D(1)\ni e^t\mape 
{\scriptsize\pmatrix{
\cos t m &{i\over M m }\sin t m \cr
i M m \sin t m &\cos t m \cr}}
\cong
{\scriptsize\pmatrix{
e^{+ti m }&0\cr
0&e^{-ti m }\cr}}\in \SO(2)
\end{eq}

In contrast
to the positive unitary time representations,
not faithful for the 
simply connected group $\D(1)$,
the free mass point is a faithful and reducible, but nondecomposable 
complex $\D(1)$-re\-pre\-sen\-ta\-tion\cite{BOE,S89} in a noncompact
indefinite unitary  group 
\begin{eq}{l} 
\D(1)\ni e^t\mape 
{\scriptsize\pmatrix{
1&{it\over M}\cr
0&1\cr}}\in\U(1,1)
\end{eq}

For the general quantum mechanical case with 
the Hamiltonian $iH=i[{\bl p^2\over 2M}+V(\bl x)]$
as basis for the represented Lie algebra $\log\D(1)\cong\R$
one obtains the time $\D(1)$-re\-pre\-sen\-ta\-tion by the ground state values 
$\angle{\com{a(s)}{b(t)}}=\angle{\com ab}(t-s)$
of the commutators with a spectral measure $\mu(m^2)$
for the time translation eigenvalues $m\in\R$ (frequencies, energies). 
In the case of a compact time development,
where there 
exists  a basis of normalizable energy eigenvectors (for the oscillator
build by the monomials of creation and annihilation operator),
the $\D(1)$-re\-pre\-sen\-ta\-tion reads
with a positive definite energy measure $\mu(m^2)\ge0$
\begin{eq}{l}
\angle{{\scriptsize\pmatrix{
\com{i\bl p}{\bl x}&\com {\bl x}{\bl x}\cr
\com {\bl p}{\bl p}&\com {\bl x}{-i\bl p}\cr}}   }(t)=
\int_0^\infty  d m^2\mu(m^2){\scriptsize\pmatrix{
\cos tm&{i\over Mm}\sin tm\cr
iMm \sin tm&\cos tm\cr}}
\end{eq}

\subsection{The Representation Defect of Particle Fields}

Particle fields are appropriate to describe {\it free} particles, i.e.
representations of the spacetime tangent structures 
leading to the particle characterization
by a causal translation property mass $ m \ne0$ or $ m =0$ with a rotation
property $\SU(2)$-spin and a $\U(1)$-polarization
resp.\cite{WIG} What about representations of the
nonlinear global spacetime model
\begin{eq}{l}
\D(2)\cong\D(\bl 1_2)\x\SO^+(1,3)/\SO(3)
\end{eq}which contains  the 
rotation classes of the Lorentz transformations 
in addition to the causal  group?

An appropriate example is a Dirac field $\bl \Psi$ for a particle
with nontrivial mass $ m $, e.g. for the electron-positron,
with the quantization
\begin{eq}{ll}
\acom{\ol{\bl\Psi}}{\bl\Psi}(x)&
=\int{d^4q\over(2\pi)^3}\ep(q_0)(\ga^kq_k+ m )\de(q^2- m ^2)
 e ^{xiq}\cr
&=\ga^0\de(\rvec x)\hbox{ for }x_0=0
\end{eq}causally  supported, i.e. 
$\acom{\ol{\bl\Psi}}{\bl\Psi}(x)=0$ for $x^2<0$.

The Dirac field is decomposable in left and right handed part with
the Weyl matrices $\si^k\cong(\bl 1_2,\rvec \si)\cong\d\si_k$
\begin{eq}{l}
\bl\Psi(x)=\bl l(x)\pl \bl r(x),~~ 
\ol{\bl\Psi}(x)=\bl\Psi^*(x)\ga^0=   \bl r^*(x)\pl\bl l^*(x)\cr
\end{eq}The field quantization
\begin{eq}{rl}
\ga_0\acom{\ol{\bl\Psi}}{\bl\Psi}(x)={\scriptsize\pmatrix{
\acom{ \bl l^*}{\bl l}&\acom{ \bl r^*}{\bl l}\cr
\acom{ \bl l^*}{\bl r}&\acom{ \bl r^*}{\bl r}\cr}}(x)
&=
\int{d^4q\over(2\pi)^3}\ep(q_0)
{\scriptsize\pmatrix{q_k\d\si_0\si^k&m\d\si_0\cr
 m\si_0&q_k\si_0\d\si^k\cr}}\de(q^2- m ^2) e ^{xiq}\cr
\p_k\d\si^k \bl l(x)=im\bl r(x),&\p_k\si^k \bl r(x)= im \bl l(x)\cr
\end{eq}has to be compared with the energy spectral representation
of the quantum mechanical time representation for the harmonic oscillator
\begin{eq}{rl}
{\scriptsize\pmatrix{
\com{i\bl p}{\bl x}&\com{\bl x}{\bl x}\cr
\com{\bl p}{\bl p}&\com {\bl x}{-i\bl p}\cr}}(t)&=
\int dE\ep(E)
{\scriptsize\pmatrix{E&{1\over M}\cr
M m ^2&E\cr}}\de(E^2- m ^2) e ^{tiE}\cr
&={\scriptsize\pmatrix{\cos tm&{i\over Mm}\sin tm\cr iMm\sin tm&\cos tm\cr}}
=\bl1_2\hbox{ for }t=0\cr
{d\over dt}\bl x(t)={1\over M}\bl p(t),&{d\over dt}\bl p(t)=-M m ^2\bl x(t)
\end{eq}

Particle fields give a causally supported 
position space  distribution of a time group
$\D(1)$-re\-pre\-sen\-ta\-tion as seen in the position space integral 
(time projection) of the
quantization condition for a Dirac particle field
\begin{eq}{ll}
\int d^3 x\ga_0\acom{\ol{\bl\Psi}}{\bl\Psi}(x)&=\int d^3 x{\scriptsize\pmatrix{
\acom{ \bl l^*}{\bl l}&\acom{ \bl r^*}{\bl l}\cr
\acom{ \bl l^*}{\bl r}&\acom{ \bl r^*}{\bl r}\cr}}(x)\cr
&=
\int dE~\ep(E)
{\scriptsize\pmatrix{E\bl 1_2&m\bl1_2\cr
 m \bl1_2&E\bl 1_2\cr}}\de(E^2- m ^2) e ^{x_0 iE}\cr
&={\scriptsize\pmatrix{
\cos x_0 m ~\bl1_2&i\sin x_0 m ~\bl1_2\cr
i  \sin x_0 m ~\bl1_2&\cos x_0 m ~\bl1_2\cr}}
\end{eq}where the momentum and position space integrations have been
interchanged.
For every time $x_0$
the position space integration goes over a compact
sphere $\{\rvec x\mid \rvec x^2\le x_0^2\}$.

The integration with respect to the time translations 
displays the Yukawa interaction and force
\begin{eq}{ll}
2\pi\int d x_0\ep(x_0)\ga_0
\acom{\ol{\bl\Psi}}{\bl\Psi}(x)&=
2\pi\int d x_0\ep(x_0)
{\scriptsize\pmatrix{
\acom{ \bl l^*}{\bl l}&\acom{ \bl r^*}{\bl l}\cr
\acom{ \bl l^*}{\bl r}&\acom{ \bl r^*}{\bl r}\cr}}(x)\cr
&=\int dQ
{\scriptsize\pmatrix{|Q|{\rvec\si\rvec x\over|\rvec x|}&im\bl 1_2\cr
im\bl 1_2&-|Q|{\rvec\si\rvec x\over|\rvec x|}\cr}}\vth(Q^2- m ^2)
 e ^{-|\rvec xQ|}\cr
&=
{\scriptsize\pmatrix{
{1+|\rvec x m |\over|\rvec x|}{\rvec\si\rvec x\over|\rvec x|}
&im\bl 1_2\cr
 im\bl 1_2 
&-{1+|\rvec x m |\over|\rvec x|}{\rvec\si\rvec x\over|\rvec x|}\cr}}
{ e ^{-|\rvec x m |}\over|\rvec x|}
\end{eq}

The rank 1 homogeneous  boost manifold $\SO^+(1,3)/\SO(3)$ contains 
as maximal abelian subgroup
the Lorentz transformations $\SO^+(1,1)$ isomorphic to  
a dilatation group $\D(1)$ with representations characterized by a mass
(inverse length) $m$
\begin{eq}{l}
\D(1)\ni e ^{x}\mape 
{\scriptsize\pmatrix{
\cosh x m   &\sinh x m  \cr
\sinh x m  &\cosh x m  \cr}}\cong{\scriptsize\pmatrix{
 e ^{+x m } &0\cr
0& e ^{- x m }\cr}}\in\SO^+(1,1)
\end{eq}

Particle fields involve representations
only for the time group $ e ^{x_0}\in\D(1)$, but not 
for the abelian boost group $ e ^{\pm |\rvec x|}\in \SO^+(1,1)$ 
as seen in the quantization of the left handed  Weyl field $\bl l(x)$
\begin{eq}{rll}
{1\over2}\tr\int d^3 x\acom{\bl l^*}{\bl l}(x)
&\hskip-2mm=\int dE\ep(E)E\de(E^2- m ^2)~ e ^{x_0 iE}&\hskip-2mm=\cos
x_0 m \cr
\pi\tr{\rvec\si\rvec x\over|\rvec x|}\int dx_0\ep(x_0)\acom{\bl l^*}{\bl l}(x)
&\hskip-2mm=\int dQ\ep(Q)Q\vth(Q^2- m ^2)
 e ^{-|\rvec xQ|}&\hskip-2mm=
{1+|\rvec x m |\over\rvec x^2}
 e ^{-|\rvec x m |}~\cr
\end{eq}$ e ^{-|\rvec xQ|}$ in the integrand 
is a  matrix element for the representation 
of the boost 
group $\SO^+(1,1)\cong \D(1)$. 
The well known Yukawa  
singularity structure ${1\over|\rvec x|}$, ${1\over\rvec x^2}$ arising 
after integration 
with the spectral functions $\vth(Q^2-m^2)$, $Q^2\vth(Q^2-m^2)$ 
for  the tangent appropriate particle fields 
cannot occur in $\SO^+(1,1)$ representations. 
A quantum representation of the spacetime model $\D(2)$ cannot be
achieved alone by particle fields, 
genuine spacetime nonparticle field contributions have to occur.

\subsection{Residual Representations for $\U(1)$ and $\D(1)$}

The Lie algebra 
$\log\U(n)\cong i\R(n)$ and the spacetime translations $\R(n)$  
are unitarily diagonalizable with $n$ Cartan coordinates
in the polar decomposition 
\begin{eq}{l}
\R(n)\cong\R^n\x\SU(n)/\U(1)^{n-1}:~~\left\{\begin{array}{rl}
i\al&=u(\al)\o i\diag \al\o u(\al)^\star\cr
x&=u(x)\o\diag x\o u(x)^\star\cr 
 \end{array}\right.\cr
\hbox{e.g. }n=2:~
i\diag \al={\scriptsize\pmatrix{
i(\al_0+|\rvec \al|)&0\cr
0&i(\al_0-|\rvec \al|)\cr}},~~\diag x={\scriptsize\pmatrix{
x_0+|\rvec x|&0\cr
0&x_0-|\rvec x|\cr}}\cr
\end{eq}leading to the 
Lie group and spacetime manifold as  manifold products
 of a maximal abelian Cartan
subgroup and a compact submanifold 
\begin{eq}{lll}
\U(n)\cong\U(1)^n\x\SU(n)/\U(1)^{n-1}:&
  e ^{i\al}&=u(\al)\o   e ^{i\diag \al}\o u(\al)^\star\cr
\D(n)\cong\D(1)^n\x\SU(n)/\U(1)^{n-1}:&
  e ^x&=u(x)\o   e ^{\diag x}\o u(x)^\star\cr 
\end{eq}Corresponding manifold products hold for the
boost structure and the simple Lie symmetry
\begin{eq}{ll}
\SU(n)\cong\U(1)^{n-1}\x\SU(n)/\U(1)^{n-1}:&\tr \al=0\cr
\SD(n)\cong\D(1)^{n-1}\x\SU(n)/\U(1)^{n-1}:&\tr x=0\cr
\end{eq}

The  Cartan subsymmetry  
for the  compact groups $\SU(n)$ and $\U(n)$ given by $\U(1)$-powers
(tori) has its analogue
in the $\D(1)$-powers 
(planes) as noncompact Cartan 
subsymmetry for the boost and causal symmetric spaces
$\SD(n)$ and $\D(n)$.

The unitary irreducible  representations of 
the abelian group $\GL(\C_\R)=\D(1)\x \U(1)$,
necessarily 1-dimensional,
have to be in $\U(1)$ since there is only one unitarity type in $\GL(\C_\R)$.
They must have an imaginary weight for the noncompact group
$\D(1)\cong\R$ and an integer winding number for the 
periodic phase group $\U(1)\cong\R/\Z$ 
\begin{eq}{rcl}
\D(1)\x\U(1)&\map&\U(1)\subnoteq\GL(\C_\R)\cr
  e ^{t+i\al}&\mape& 
  e ^{t\de+i\al z} 
\then \left\{
\begin{array}{l}
\ol\de=-\de=i m \in i\R\cr
 z\in\Z\end{array}\right.\cr
\end{eq}which leads to 
the representation weights, identical with the invariants 
\begin{eq}{l} 
\weights\GL(\C_\R)=\weights\D(1)\x\weights\U(1)=\{(im,z)\}=i\R\x\Z
\end{eq}

An irreducible representation of the complex group $\GL(\C)$ 
arises as residue of its eigenvalue as  singularity 
by using the 
complex Lie algebra forms $Q\in\C$ 
\begin{eq}{l}
\GL(\C)\ni   e ^Z\mape   e ^{Z\ze}={1\over2i\pi}
\oint dQ{1\over Q-\ze}  e ^{ZQ},~~\ze\in\C
\end{eq}which gives for  the
unitary irreducible $\U(1)$ and 
$\D(1)$-re\-pre\-sen\-ta\-tions
\begin{eq}{rlll}
\U(1)\ni   e ^{i\al}&\mape   e ^{i\al z}&=
{1\over 2i\pi}\oint dw{1\over w-z}  e ^{i \al w},&z\in \Z\cr
\D(1)\ni   e ^{t}&\mape   e ^{t i m}&=
{1\over 2i\pi}\oint dq{ 1\over q- m}  e ^{t iq},& im\in i\R\cr
\end{eq}The integration for the 
noncompact and compact group are related 
to each other for the
Lie algebras and their  forms
\begin{eq}{c}
\hbox{for }\GL(\C)~~(Z,Q),~~Z=t+i\al,~Q=q+iw\cr
\hbox{for }\D(1)~~(t,q)\lrmap (i\al,iw)~~\hbox{for }\U(1)
\end{eq}The  nontrivial irreducible representations of $\U(1)$ and $\D(1)$ are
not selfdual.

{\it Measured representations} use  measures of the weights.
The integer weights for the compact group $\U(1)$
have  as discrete complex measures series 
of complex numbers leading to Fourier series  
\begin{eq}{rl}
\meas\Z\ni \{\mu_z\}_{z\in\Z}&\mape\rep\U(1),~~\mu_z\in\C\cr
\U(1)\ni   e ^{i\al}&\mape
{\SUM_{z\in\Z}}\mu_z  e ^{i\al z}\cr
\end{eq}The continuous weights  for $\D(1)$ have Lebesque measure $dm$ based
complex measures giving  rise to Fourier integrals 
\begin{eq}{rl}
\meas\R\ni\mu&\mape\rep\D(1)\cr
\D(1)\ni   e ^{t}&\mape
\int dm\mu(m)  e ^{t i m}\cr
\end{eq}

The unitary  
irreducible  representations of  the simple group $\SL(\C^2_\R)$
are characterized by selfdual re\-pre\-sen\-ta\-tions of 
a  Cartan subgroup 
\begin{eq}{l}
\GL(\C_\R)\si^3=\D(1)\si^3\x\U(1)\si^3
\cong\SO^+(1,1)\x\SO(2)
\end{eq}which can go
in the two types of 2-di\-men\-sio\-nal unitary groups,
the definite unitary $\SU(2)$ or the indefinite unitary  $\SU(1,1)$
\begin{eq}{rcl}
\SL(\C^2_\R)\supnoteq \D(1)\si^3\x\U(1)\si^3&\map&
\left\{
\begin{array}{c}
\U(1)\si^3\subnoteq\SU(2)\cr\D(1)\si^3\subnoteq\SU(1,1)\cr\end{array}\right\}
\subnoteq\SL(\C^2_\R)\cr
  e ^{( x_3 +i\al_3)\si^3}
&\mape&   e ^{( x_3\de_3+i\al_3 z_3)\si^3}\cr
\end{eq}This defines the 
weights $(\de_3,z_3)$ of the principal and supplementary 
series
for $\SU(2)$ and $\SU(1,1)$ resp.
\begin{eq}{lll}
\weights^{(2,0)}\SL(\C^2_\R)&=\{(i m _3,z_3)\}&=i\R\x \Z=\weights\GL(\C_\R)\cr
\weights^{(1,1)}\SL(\C^2_\R)&=\{(m _3,0)\}&=\R\cr
\end{eq}

The principal series $\GL(\C_\R)\si^3$-weights coincide with the 
$\GL(\C_\R)$-weights.
One $\GL(\C_\R)\si^3$-re\-pre\-sen\-ta\-tion is characterized by a dual pair
 $\{\pm im_3\}$ for $\D(1)\si^3$ and
$\{\pm z_3\}$ for $\U(1)\si^3$.
The new real $\D(1)\si^3$-weights $m_3\in\R$
in contrast to the imaginary $\D(1)$-weights $im\in i\R$ 
above are possible for dimensions
$n\ge2$ with the possibility of  
indefinite unitary groups. One $\SO^+(1,1)$-re\-pre\-sen\-ta\-tion  in $\SU(1,1)$
is characterized by a dual pair $\{\pm m_3\}$.
For  dimensions $n\ge3$ no additional  types of invariants 
arise for 
the  representations of the Cartan subgroups $\U(1)$ and $\D(1)$.
Altogether the unitary 
$\U(1)$ and $\D(1)$-re\-pre\-sen\-ta\-tions
are characterizable by the   invariants
\begin{eq}{rclccccl}
\irrep\U(1)&\pl&\irrep\SO(2)&\cong&\{z\}&\pl&\{2J\}&=\Z\pl\N\cr
\irrep\D(1)&\pl &\irrep\SO^+(1,1)&\cong&\{im\}&\pl&\{-m^2\}&
= i\R\pl \R^-\cr
\end{eq}Generalized functions have to be given
taking care of the quadratic invariants 
as complex plane singularities for selfdual 
residual representations.

Pairs of dual irreducible
$\U(1)$-re\-pre\-sen\-ta\-tions 
$\{  e ^{\pm i\al m}\mid m\in\Z\}$ can be formulated by
measures with 
the integration prescription  $m^2\pm io=(|m|\pm io)^2$ for the invariant
\begin{eq}{l}
  e ^{\pm i|\al m|}=
\pm {1\over i\pi}\int dw{|m|\over w^2\mp io -m^2}  e ^{i\al w},~~m\in\R\cr
\end{eq}If the Cartan subgroup $\U(1)$  comes in the special group 
$\SU(n)$, $n\ge2$, the  residual representation
employs the  forms of the $\R^{n^2-1}$-dimensional 
tangent Lie algebra 
with the singularity of the generalized functions
determined by the values of the  invariant multilinear forms,
starting for $n=2$ with the bilinear Killing form $\rvec q^2$
and a dipole  
\begin{eq}{c}
\hbox{for }
\U(1)\si^3\cong\SO(2):~  e ^{\pm i|\rvec \al m|}
=\pm{1\over i\pi^2}
\int d^3 w{|m|\over (\rvec w^2-m^2\mp io)^2}  e ^{i\rvec \al\rvec w},~~m\in\R
\cr
 \irrep\SO(2)\cong \{|m|=2J\}=\N\cr
\end{eq}

Pairs of  dual irreducible $\D(1)$-re\-pre\-sen\-ta\-tions
$\{  e ^{\pm xm}\mid m\in\R\}$ are obtained by $(i\al,iw)\lrmap(x,q)$
leading to  the following   Lie algebra form measure
\begin{eq}{l}
  e ^{-|xm|}
={1\over \pi}\int dq{|m|\over q^2+m^2}  e ^{-xiq},~~m\in\R\cr
\end{eq}For a boost manifold $\SD(n)$, $n\ge2$, 
the $\D(1)\si^3$-re\-pre\-sen\-ta\-tions use the $\R^{n^2-1}$-dimensional
tangent space forms  (momenta), again with a dipole for $n=2$
\begin{eq}{c}
 \hbox{for }\D(1)\si^3\cong\SO^+(1,1):~~  e ^{-|\rvec x m|}
={1\over \pi^2}
\int d^3 q{|m|\over (\rvec q^2+m^2)^2}  e ^{-\rvec x i\rvec q},~~m\in\R
\cr
\irrep\SO^+(1,1)\cong \{-m^2\}=\R^-\cr
\end{eq}

\subsection{Residual Representations for Spin $\SU(2)$}

The matrix elements of the irreducible  $\SU(2)$-re\-pre\-sen\-ta\-tions
$[2J]$ by unitary
$\C^{2J+1}$-automorphisms can be given via measures of the Lie algebra forms
supported by integers.

With the generalized function singularities 
as  angular momenta values 
\begin{eq}{rl}
\SU(2)&\cong\SU(2)/\U(1)\x\U(1)\si^3\cr
&\cong\SO(3)/\SO(2)\x\SO(2)\cr
[\pm 1](\rvec\al)
&=
{1\over \pi^2}
\int d^3w{\rvec w\over (\rvec w^2-1\mp io)^2}
  e ^{i\rvec \al\rvec w}
=i{\rvec \al\over|\rvec \al|}  e ^{\pm i|\rvec \al|}
\cr
\end{eq}there arise  the matrix elements of the
{\it fundamental Pauli re\-pre\-sen\-ta\-tion}
\begin{eq}{l}
  e ^{i\rvec\al\rvec \si}
=\bl1_2\cos|\rvec\al|
+i{\rvec\si\rvec\al\over|\rvec\al|}\sin|\rvec\al|
\end{eq}

Using the irreducible $\SO(3)$-polynomials $[\rvec w]^{2J}$,
homogeneous of degree $2J$ in the
angular momenta
\begin{eq}{l}
[\rvec w]^0=1,~~
[\rvec w]^1=\{w_a\mid a=1,2,3\},~~
[\rvec w]^2=\{w_aw_b-{\de_{ab}\over 3}\rvec w^2\},~~\dots
\end{eq}the residual formulation for the matrix elements
of the nontrivial {\it 
irreducible $\SU(2)$-re\-pre\-sen\-ta\-tion}   reads
\begin{eq}{rl}
\SU(2)\ni  e ^{i\rvec\al\rvec\si}\mape[\pm2J](\rvec\al)=&
{1\over \pi^2}
\int d^3w{[\rvec w]^{2J} \over (\rvec w^2-4J^2\mp io)^{2+J-c(J)}}
  e ^{i\rvec \al\rvec w}\cr
&2J=1,2\dots,~~\cr
\end{eq}The $\SU(2)$-centrality (two-ality) $2c(J)$  
is trivial for integer $J$ and $1$ for
halfinteger $J$
\begin{eq}{l}
2c(J)=\left\{\begin{array}{rl}
0,&2J=0,2,4,\dots\cr
1,&2J=1,3,\dots\cr\end{array}\right.
\end{eq}

All representation elements of $\SU(2)$ can be obtained by
derivations with respect 
to the invariant $m^2$  and the Lie parameter $\rvec\al$
from the {\it Yukawa potential for $\SU(2)$}, 
defined in analogy to the
usual Yukawa potential (next section), which is no 
$\SU(2)$-re\-pre\-sen\-ta\-tion because of the Lie parameter $\rvec \al=0$
singularity
\begin{eq}{rll}
{1\over \pi^2}
\int d^3w{1\over \rvec w^2-m^2\mp io}  e ^{i\rvec \al\rvec w}
&=2{  e ^{\pm i|\rvec \al m|}\over|\rvec\al|},&m\in\R,~\rvec\al\ne0\cr
{\p\over \p m^2}={1\over 2|m|}{\p\over\p|m|},&
{\p\over\p\rvec\al}={\rvec\al\over|\rvec\al|}{\p\over\p|\rvec\al|}&\cr
\end{eq}The $m^2$-derivative leads to
\begin{eq}{l}
{1\over \pi^2}
\int d^3w{1\over (\rvec w^2-m^2\mp io)^2}  e ^{i\rvec \al\rvec w}
=\pm i{  e ^{\pm i|\rvec \al m|}\over|m|},~~m\in\R,~m\ne0\cr
\end{eq}which gives
the trivial representation  $[0](\rvec\al)=1$ for an appropriate limit $m\to 0$.

The representation matrix elements come  in  a product of 
a $\U(1)\si^3$-re\-pre\-sen\-ta\-tion factor 
with the invariant $2J$ (rotation frequency) multiplying
the modulus of the Lie parameter $|\rvec\al|$ 
and a polynomial  in the Lie parameter direction
 ${\rvec\al\over|\rvec\al|}$ (rotation axis), homogeneous of
degree $2J$, representing the 2-dimensional  
symmetric space ($2$-sphere) $\SU(2)/\U(1)\cong \SO(3)/\SO(2)$
\begin{eq}{l}
[\pm2J](\rvec\al)\sim
|\rvec\al |^{1-2c(J)}[{i\rvec\al\over|\rvec\al|}]^{2J}
   e ^{\pm i 2J|\rvec \al|}
\end{eq}e.g. the adjoint representation
\begin{eq}{l}
\SO(3):~~[\pm 2](\rvec\al)
={1\over \pi^2}
\int d^3w{w_aw_b-{\de_{ab}\over 3}\rvec w^2 \over 
(\rvec w^2-4\mp io)^3}
  e ^{i\rvec \al\rvec w}
= -{|\rvec\al|\over4}\({\al_a\al_b\over\rvec\al^2}-{\de_{ab}\over3}\)
  e ^{\pm 2i|\rvec \al|}
\cr
\end{eq}to be compared with the elements in the $(3\x3)$-matrix
\begin{eq}{l}
\de_{ab}\cos2|\rvec \al|+{\al_a\al_b\over\rvec\al^2}(1-
\cos2|\rvec \al|)+\ep_{abc}{\al_c\over|\rvec \al|}\sin 2|\rvec \al|
\end{eq}

A  measured $\SU(2)$-re\-pre\-sen\-ta\-tion 
is a  
Fourier series as for $\U(1)$
where each term comes
with a unique $\SU(2)/\U(1)$-polynomial 
\begin{eq}{rl}
\meas\Z\ni\{\mu_z\}_{z\in\Z}&\mape \rep\SU(2),~~\mu_z\in\C\cr
\SU(2)\ni  e ^{i\rvec\al\rvec\si}&\mape
{\SUM_{2J=0,1,\dots}}\(\mu_{2J}[2J]+\mu_{-2J}[-2J]\)\cr
\end{eq}

\subsection{Residual Representations for Boost $\SD(2)$}

The unitary representations of the 
globally symmetric space $\SL(\C^2_\R)/\SU(2)$,
called boost manifold $\SD(2)$,
will be defined via the polar decomposition in  a noncompact Cartan group
$\D(1)$, in contrast to the
compact $\U(1)$ for $\SU(2)$, and a compact submanifold
$\SU(2)/\U(1)$, identical for $\SU(2)$ and $\SD(2)$
\begin{eq}{rll}
\SD(2)&=\SL(\C^2_\R)/\SU(2)&\cong\SU(2)/\U(1)\x\D(1)\si^3\cr
&\cong\SO^+(1,3)/\SO(3)&\cong\SO(3)/\SO(2)\x\SO^+(1,1)\cr
\end{eq}by using the tangent space relations
\begin{eq}{l}
\hbox{for }\SD(2)~~(\rvec x,\rvec q)\lrmap (i\rvec \al,i\rvec w)~~
\hbox{for }\SU(2)
\end{eq}

With the momenta measure singularities
 for the 
tangent space forms at
the representation invariant $-m^2$
one obtains the {\it fundamental $\SD(2)$-re\-pre\-sen\-ta\-tions} 
\begin{eq}{l}
[m^2;1](\rvec x)
={1\over \pi^2}
\int d^3 q{i\rvec q\over (\rvec q^2+m^2)^2}  e ^{-\rvec x i\rvec q}
={\rvec x\over|\rvec x|}  e ^{-|\rvec x m|},~~m\in\R\cr
\end{eq}to be compared with
\begin{eq}{l}
  e ^{\rvec x |m|\rvec \si}
=\bl1_2\cosh|\rvec x m|
+{\rvec\si\rvec x\over|\rvec x|}\sinh|\rvec xm|
\end{eq}and, in general, with the $\SO(3)$-irreducible 
momenta polynomials $[\rvec q]^{2J}$,
the {\it irreducible
$\SD(2)$-re\-pre\-sen\-ta\-tions} 
\begin{eq}{rl}
\SD(2)\ni  e ^{\rvec x\rvec\si}\mape[m^2;2J](\rvec x)=&
{1\over \pi^2}
\int d^3 q{[i\rvec q]^{2J} \over (\rvec q^2+m^2)^{2+J-c(J)}}
  e ^{-\rvec x i\rvec q}\cr
&m\in\R,~2J=0,1,2,\dots\cr
\end{eq}

In contrast to the group $\SU(2)$ where the representations
of the compact factors $\U(1)\si^3$
and $\SU(2)/\U(1)$ have to be related to each other 
by the invariant $2J$,
in the symmetric space
$\SL(\C^2_\R)/\SU(2)$ 
the invariant $m^2$ of the noncompact
Cartan group $\D(1)\si^3$-re\-pre\-sen\-ta\-tion is not related 
to the degree $2J$ of the homogenous polynomial for the representation
of the compact sphere $\SO(3)/\SO(2)$.

All $\SD(2)$-re\-pre\-sen\-ta\-tions  can be obtained by
derivations ${\p\over\p m^2}$ and ${\p\over\p\rvec x}$ 
from the {\it Yukawa potential} which, by itself, is no $\SD(2)$-re\-pre\-sen\-ta\-tion
because of the $\rvec x=0$ singularity
\begin{eq}{rl}
{1\over \pi^2}
\int d^3 q{1\over \rvec q^2+m^2}  e ^{-\rvec x i\rvec q}
&=2{  e ^{-|\rvec x m|}\over |\rvec x|}
,~~m\in\R,~\rvec x\ne0\cr
\end{eq}The {\it scalar representations}
\begin{eq}{l}
[m^2;0](\rvec x)
={1\over \pi^2}
\int d^3 q{1\over (\rvec q^2+m^2)^2}  e ^{-\rvec x i\rvec q}
={  e ^{-|\rvec x m|}\over |m|}
,~~m\in\R,~m\ne0\cr
\end{eq}are  trivial for the sphere $\SO(3)/\SO(2)$.
The analogue to the adjoint spin representation reads
\begin{eq}{l}
[m^2; 2](\rvec x)
={1\over \pi^2}
\int d^3 q{-q_aq_b+{\de_{ab}\over 3}\rvec q^2 \over 
(\rvec q^2+m^2)^3}
  e ^{-\rvec xi\rvec q}
={|\rvec x |\over4} \({x_ax_b\over\rvec x^2}-{\de_{ab}\over3}\)
  e ^{-|\rvec x m|}
\end{eq}All irreducible representations
 can be written as products 
 \begin{eq}{l}
[m^2;2J](\rvec x)
\sim  |\rvec x |^{1-2c(J)}[{\rvec x\over|\rvec x|}]^{2J}
   e ^{-|\rvec x m|}\cr
\end{eq}

A measured $\SD(2)$-re\-pre\-sen\-ta\-tion 
is a sum over the spin numbers $2J$ with measures $\mu_{2J}$ for the 
continuous invariants
\begin{eq}{rl}
\meas\N\x\R^+&\ni\{\mu_{2J}\}_{2J\in\N}\mape \rep\SD(2)\cr
\SD(2)\ni  e ^{\rvec x\rvec\si}
&\mape{\SUM_{2J=0,1,\dots}}
\int_0^\infty  dm^2\mu_{2J}(m^2) [m^2;2J](\rvec x)\cr
&={\SUM_{2J=0,1,\dots}}
\int_0^\infty  dm^2\mu_{2J}(m^2) 
{1\over \pi^2}
\int d^3 q{[i\rvec q]^{2J} \over (\rvec q^2+m^2)^{2+J-c(J)}}
  e ^{-\rvec x i\rvec q}\cr
\end{eq}

The two integrations in measured representations go over the tangent space forms 
$\int d^3q$
and the invariants $\int_0^\infty dm^2$
with the dimensions $3$ and $1$ of 
the symmetric space and a Cartan subgroup resp. 
For the measured $\SU(2)$-re\-pre\-sen\-ta\-tion 
in the former section the 1-dimensional
integration is replaced by a discrete sum.

\subsection{Two Continuous Invariants for Spacetime}

Since Yukawa, the unification of a time development, characterized by a
particle mass $ |m _0|$, with a position space  interaction,
characterized by a  range ${1\over |m _3|}$,
in one spacetime Klein-Gordon equation with one mass
\begin{eq}{l}
\left.\begin{array}{rl}
({d^2\over dt^2}+ m_0^2){  e ^{i|tm_0|}\over 2i|m_0|}&=\de(t)\cr
(-{\p^2\over\p\rvec x^2}+ m_3^2){  e ^{-|\rvec x m_3|}\over4\pi|\rvec x|}&=
\de(\rvec x)\cr\end{array}\right\}
\then \begin{array}{l}
(\p^2+ m ^2)G(x)=\de(x)\cr
\hbox{with } m_0^2= m_3^2= m ^2\end{array}
\end{eq}seems to be an obvious relativistic bonus.

Particle fields with a Dirac energy-mo\-men\-tum measure 
in their quantization
\begin{eq}{l}
\bl c_j(x|m_0)=\int {d^4q\over(2\pi)^3}\ep(q_0)q_j\de(q^2- m_0^2)
  e ^{xiq}\cr
\end{eq}give by  position space integration 
a Dirac measure for the time weights $iq_0\in i\R$ (real energies $q_0$),
selfdually supported at $\pm i m_0$, leading to
$\SO(2)$-re\-pre\-sen\-ta\-tion matrix elements of the 
abelian time group $\D(1)$
\begin{eq}{l}
\int d^3 x \bl c_j(x|m_0)
 =\de_j^0\int d^1q\ep(q)q \de(q^2- m_0^2)~  e ^{ x_0 iq}
 =\de_j^0\cos x_0 m_0\cr
\end{eq}

The appropriate  measure for a representation 
of the boost subgroup $\D(1)\si^3\cong\SO^+(1,1)$  arises from
a derived energy-mo\-men\-tum Dirac measure 
\begin{eq}{l}
\bl c^{\rm dip}_j(x|m_3)=-{ d\bl c(x|m_3)
\over dm_3^2}=\int {d^4q\over(2\pi)^3}\ep(q_0)q_j\de'(q^2-m_3^2)
  e ^{xiq}\cr
\end{eq}Time integration leads to
a Dirac measure for the $\SO^+(1,1)$-invariant
and an $\SO^+(1,1)$-re\-pre\-sen\-ta\-tion
\begin{eq}{l}
4\pi{x_a\over|\rvec x|}\de^a_j\int dx_0\ep(x_0)
\bl c_j^{\rm dip}(x|m_3)
=2\int d^1q\ep(q)q\de(q^2- m_3^2)~  e ^{-|\rvec xq|} 
=  e ^{-|\rvec x m_3|}\cr
\end{eq}

The appropriateness of the Dirac energy-momentum measure 
for time in contrast to the  derived  measure 
for position space 
\begin{eq}{l}
(-{\p^2\over\p\rvec x^2}+ m_3^2)^2
{  e ^{-|\rvec x m_3|}\over 8\pi |m_3|}=\de(\rvec x)
\end{eq}reflects the different dimensions 1 and 3 resp. as seen 
also in the
 energy-momentum Lebesque measure $d^4 q= dq_0~\rvec q^2d|\rvec q|~
 d\phi d\!\cos\th$.

The association of the singularities at $m_0^2$ and $m_3^2$
to  representation invariants for $\D(1)$ (time) and $\SO^+(1,1)$
resp. is blurred since a tangent space decomposition
$x=\bl1_2x_0+\rvec\si\rvec x$ 
into time and position space translations is not compatible with the
action of the Lorentz
group $\SO^+(1,3)$. The Dirac measure has also a nontrivial 
projection for the boost $\SO^+(1,1)$ and the derived Dirac measure 
a nontrivial projection for time $\D(1)$
\begin{eq}{rl}
4\pi{ x_a\over|\rvec x|}\de^a_j\int dx_0\ep(x_0)
\bl c_j(x|m_0)
&=2{1+|\rvec x m_0|\over \rvec x^2}  e ^{-|\rvec x m_0|}\cr
\int d^3 x 
\bl c_j^{\rm dip}(x|m_3)
&=\de_j^0{x_0\sin x_0 m_3\over 2m_3}\cr
\end{eq}The $\D(1)$-projection of the derived Dirac measure leads to matrix
elements of reducible nondecomposable time representations\cite{S89}.
The boost projection of the  Dirac measure leads to 
a Yukawa force which is not related to a  matrix
element of an $\SO^+(1,1)$-re\-pre\-sen\-ta\-tion.

An ordered integration $d^4q\ep(q_0)$
with an energy-mo\-men\-tum
Dirac  measure coincides with 
an integration with an energy-mo\-men\-tum
principal value $\ro P$  pole measure as shown by the identities
\begin{eq}{l}
\int d^4q ~\ep(x_0q_0)\de^{(N)}( m ^2-q^2)
  e ^{xiq}=
{1\over i\pi}\int d^4q {\Ga(1+N)
\over (q_\ro P^2- m ^2)^{1+N}}  e ^{xiq},~~
N=0,1,\dots
\end{eq}

Related 
to  two Cartan coordinates $x_0\pm|\rvec x|$ 
which reflect the
real rank 2 of the noncompact homogeneous manifold 
$\D(2)=\GL(\C^2_\R)/\U(2)$,
i.e. two abelian subgroups $\D(\bl1_2)$ (time) and
$\SO^+(1,1)$ as a subgroup of the boost manifold $\SO^+(1,3)/\SO(3)$,
two invariants are appropriate as support
for the measures of the energy-momentum space with the action of the Lorentz
group.

The unitary  
irreducible  representations of  the
dilatation Lorentz group
\begin{eq}{rl}
\GL(\C^2_\R)/\U(\bl1_2)\cong\D(\bl 1_2)\x\SO^+(1,3)\cr
\end{eq}with Cartan subgroup $\D(\bl1_2)\x\SO^+(1,1)\x\SO(2)$
are characterized by two invariants (masses) from a continuous spectrum
for the  noncompact group $\D(\bl1_2)\x\D(1)\si^3$ (time and boost) and  one 
possibly trivial  integer invariant (winding number) 
for the compact polarization  group $\U(1)\si^3$
\begin{eq}{rcl}
\GL(\C^2_\R)/\U(\bl1_2)\supnoteq \D(\bl1_2)\x\D(1)\si^3\x\U(1)\si^3&\map&\left\{
\begin{array}{c}
\U(2)\cr\U(1,1)\cr\end{array}\right\}
\subnoteq\GL(\C^2_\R)\cr
  e ^{x_0\bl 1_2+(x_3+i\al_3)\si^3}
&\mape& 
  e ^{x_0\de_0\bl 1_2+( x_3\de_3+i\al_3z_3)\si^3}
\end{eq}leading to the weights $(\de_0,\de_3,z_3)$
for principal and supplementary series 
\begin{eq}{lll}
\weights^{(2,0)}\GL(\C^2_\R)/\U(\bl1_2)
&=\{(i m _0,i m _3,z_3)\}&=i\R\x i\R\x \Z\cr
\weights^{(1,1)}\GL(\C^2_\R)/\U(\bl1_2)
&=\{(i m _0, m _3)\}&=i\R\x\R \cr
\end{eq}The weights $(im_0,m_3)$ 
  of the supplementary series 
  with trivial $\SU(2)$-re\-pre\-sen\-ta\-tion are relevant for
representations of spacetime $\D(2)$ as the unitary classes
$\D(\bl1_2)\x\SL(\C^2_\R)/\SU(2)$.  $m_0$ characterizes
 the positive unitary 
 representations $\D(1)\ni e^{x_0}\mape e^{x_0 im_0}\in\U(1)$  
 with a  particle mass $m_0$ and a 
 probability interpretation. $m^2_3$ characterizes the indefinite
 unitary representation $\SO^+(1,1)\ni e^{-|\rvec x| }\mape
 e^{-|\rvec x m_3|}\in\SU(1,1)$ with an interaction range ${1\over |m_3|}$
and without particle asymptotics. 
There is no group theoretical reason
to identify both scales $ m^2 _0= m_3^2$ - in general, 
the representations of spacetime $\D(2)$ come with two different
scales whose ratio 
${ m_3^2\over  m_0^2}$ is a representation characteristic 
physically important constant. 

The ratio of the characterizing invariants 
should be seen  is analogy to the relative normalization
 of time and position space translations
${\scriptsize\pmatrix{{\ell^2\over c^2}&0\cr 0&-\ell^2\bl1_3\cr}}$
as given with the maximal action velocity (speed of light) $c^2$.

 \subsection{Pole  Measures of Energy-Momenta}

To generalize the  representations
of the abelian causal  group $\D(1)$ as residues for
energy singularities to representations of the 
 homogeneous  causal spacetime   $\D(n)$ 
one starts from the matrix elements of 
nondecomposable re\-pre\-sen\-ta\-tions
of a Cartan subgroup  $\D(1)^n$
with Cartan coordinates
$\{\xi_r\}_{r=1}^n$, given as products of residues
\begin{eq}{rl}
\D(1)^n\ni {\scriptsize\pmatrix{
e^{\xi_1}&\dots&0\cr
&\dots&\cr
0&\dots&e^{\xi_n}\cr}}
&\mape 
{(i\xi_1)^{ N _1}\cdots (i\xi_n)^{ N _n}
\over  N _1!\cdots  N _n!}e^{\xi_1 im_1+\dots+\xi_nim_n}\cr
&={1\over(2i\pi )^n}\oint d^n q~
{e^{\xi_1i q_1+\dots+\xi_n iq_n}\over
( q_1-m_1)^{1+ N _1}
\cdots ( q_n-m_n)^{1+ N _n}}\cr
& N_r=0,\dots,N_r,~~r=1,\dots,n\cr
\end{eq}with real invariants $\{m_r\}_{r=1}^n$ 
(Cartan masses) and nildimensions
$\{N_r\}_{r=1}^n$, trivial for the irreducible representations.

If the group $\D(1)^n$ comes as Cartan subgroup 
in the  spacetime  manifold  $\D(n)$ 
\begin{eq}{l}
 \D(1)^n\inmap \D(n)\cong\D(1)^n\x\SU(n)/\U(1)^{n-1}
\end{eq}one embeds in the  Lebesque measure $d^{n^2} q$ 
of the energy-mo\-men\-ta
\begin{eq}{l}
 d^n q=d^1q_1\cdots d^1q_n\hbox{ on }\R^n
\inmap d^{n^2}q \hbox{ on }\R^{n^2}
\end{eq}invariant under $\SL(\C_\R^n)$. The quotient
with the $n$th power of the $\SL(\C_\R^n)$-invariant determinant 
(volume element)
\begin{eq}{l}
{d^{n^2}q\over (q^n)^n}\hbox{ with }q^n=\det q=\left\{
\begin{array}{ll}
q,&n=1\cr
\det{\scriptsize\pmatrix{q_0+q_3&q_1-iq_2\cr q_1+iq_2&q_0-q_3
\cr}},&n=2\end{array}\right.
\end{eq}is a $\GL(\C_\R^n)$-invariant measure.

The $\D(1)^n$ eigenvalues are
implemented as invariant singularities
\begin{eq}{l}
{d^n q  \over ( q_1-m_1)\cdots( q_n-m_n)}\inmap 
{d^{n^2} q \over (q^n-m_1^n)\cdots (q^n-m_n^n)}
\end{eq}leading to
the {\it irreducible scalar pole  measures} 
 of the energy-mo\-men\-ta for $\GL(\C^n_\R)$ 
\begin{eq}{l}
{d^{n^2} q \over (q^n-m_1^n)\cdots (q^n-m_n^n)}=\left\{
\begin{array}{ll}
{d^1q\over q-m},&n=1\cr
{d^4 q \over (q^2-m_1^2)(q^2-m_2^2)},&n=2\end{array}\right.
\cr
\end{eq}Their invariance group  
is  the homogeneous   group $\SL(\C^n_\R)/\I(n)$,
i.e. $\SO^+(1,3)$ for $n=2$.

The  compact manifold $\SU(n)/\U(1)^{n-1}$ with
 $2{n\choose2}$ coordinates  
can be nontrivially represented by energy-momentum polynomials.

\subsection{Residual Representations of Spacetime}

Matrix elements of Lie group representations 
can be formulated  as residues for characterizing invariant singularities
of their Lie algebra forms.
This will be done also for  representations of the real rank 2
symmetric spacetime
$\D(2)$ using generalized functions of the 
noncompact $\R^4$-isomorphic energy-momenta $q\in\R(2)^T$
as linear forms  of the $\D(2)$ tangent spacetime translations.
Two energy-momentum  invariants $q^2$ characterize 
the action of the causal and boost subgroup of $\GL(\C^2_\R)$.

Representations of spacetime 
\begin{eq}{rll}
\D(2)&=\GL(\C^2_\R)/\U(2)&\cr
&=\D(\bl1_2)\x\SL(\C^2_\R)/\SU(2)
&\cong \D(\bl 1_2)\x\D(1)\si^3\x \SU(2)/\U(1)\cr
&\cong\D(\bl1_2)\x\SO^+(1,3)/\SO(3)
&\cong \D(\bl1_2)\x\SO^+(1,1)\x \SO(3)/\SO(2)\cr
\end{eq}will be built up by energy-momentum measures,
compatible with the action of the Lorentz group 
$\SO^+(1,3)$ on the tangent space.  
The $\GL(\C^2_\R)$-invariant measures of the
energy-mo\-men\-ta ${d^4q\over (q^2)^2}$
use the $\SL(\C^2_\R)$-invariant $2$-form 
$q^2$ in the denominator. The two invariant masses  
characterizing the representations
of a noncompact  Cartan subgroup representation
$\GL(\C^2_\R)\supnoteq \D(\bl1_2)\x\SO^+(1,1)\map\U(1)\x\SU(1,1)$
are implemented via singularities ${d^4q\over (q^2- m_0^2)(q^2- m_3^2)}$
in the {\it irreducible spacetime representations} 
\begin{eq}{rl}
\D(2)\ni   e ^x\mape
[m_0^2,m_3^2;2J](x)
&={1\over\pi^3}\int
d^4q{[q]^{2J} \over   (q_\ro P^2-m_0^2)(q_\ro P^2-m_3^2)^{1+J+c(J)}}
    e ^{xiq}\cr
&m_{0,3}\in\R,~~2J=0,1,\dots
\end{eq}The spin related factor 
\begin{eq}{l}
{[q]^{2J}\over (q_\ro P^2-m_3^2)^{J+c(J)}}
\end{eq}with the centrality $2c(J)\in\{0,1\}$ describes the Lorentz compatible embedding of the 
sphere $\SO(3)/\SO(2)$ representations  
via the irreducible energy-momentum $\SO^+(1,3)$-polynomials $[q]^{2J}$,
homogeneous of degree $2J$
\begin{eq}{llll}
[q]^0=\{1\},&[q]^1=\{q_j\mid j=0,1,2,3\},&
[q]^2=\{q_jq_k-{g_{jk}\over4}q^2\},&\dots\cr
\end{eq}acted on by Lorentz group
representations  $[2J|2J]$.  
For nontrivial $J$ the representations come with a multiple pole at $m_3^2$.
As shown below, the  spacetime representations 
depend on  $\vth(x^2)x$ which reflects the manifold isomorphy of spacetime
$\D(2)$ and the strictly positive cone $\{x\in\R^4\mid\spec x>0\}$
in the tangent Minkowski spacetime translations.  

The {\it measured spacetime representations}
\begin{eq}{l}
\meas\N\x\R^{+2} \ni\{\mu_{2J}^0\x\mu_{2J}^3\}_{2J\in\N}\mape\rep\D(2)\cr
\D(2)\ni   e ^x\mape
{\SUM_{2J=0,1,\dots}}\int_0^\infty d m^2_0 d m^2_3
\mu^0_{2J}(m^2_0)\mu^3_{2J}(m^2_3)
[m_0^2,m_3^2;2J](x)\cr
\end{eq}involve a product  measure for the continuous
invariants $(m^2_0,m^2_3)\in\R^+\x\R^+$.

The $\D(2)$-re\-pre\-sen\-ta\-tions are different from
the Lorentz compatible position space  distributions of time representations
used for the quantization of tangent space 
particle fields (K\"allen-Lehmann representations\cite{KL}), e.g. for $2J=1$ 
\begin{eq}{l}
\hbox{particle fields: }\int_0^\infty d m^2
\mu(m^2){1\over\pi^3}
\int d^4q {q_j\over q_\ro P^2-m^2}  e ^{xiq},~~
\mu(m^2)\ge0
\end{eq}with positive definite probability related 
spectral measure $\mu(m^2)$ for the invariants of the time
$\D(\bl1_2)$-re\-pre\-sen\-ta\-tions in $\U(1)$.

The representations of rank 2 spacetime $\D(2)$ 
have to be seen as the generalization 
of measured representations for the rank 1 abelian time group $\D(1)$  
\begin{eq}{l}
\D(1)\ni   e ^ t\mape  
\int d m\mu(m)
  e ^{tim}=\int d m\mu(m)
{\ep(t)\over i\pi}\int
d^1q{1\over q_\ro P-m}
  e ^{ tiq}
\end{eq}The irreducible unitary time $\D(1)$-re\-pre\-sen\-ta\-tions 
$  e ^ t\mape   e ^{tim}$
use a Dirac energy measure with one supporting energy $ m $.
All matrix elements of 
the nondecomposable $\D(1)$-re\-pre\-sen\-ta\-tions
are given by  derivatives with respect to the invariant
\begin{eq}{rl}
\D(1)\ni  e ^t\mape&(ti)^N  e ^{tim}
={\ep(t)\over i\pi}\int d^1q {\Ga(1+N)\over
(q_\ro P-m)^{1+N}}  e ^{tiq}
=({d\over dm})^N  e ^{tim}\cr&
m\in\R,~~N=0,1,\dots
\end{eq}

The spacetime analogue 
is given by the nondecomposable  $\D(2)$-re\-pre\-sen\-ta\-tion matrix elements 
with  two supporting masses
\begin{eq}{l}
\begin{array}{rl}
\D(2)\ni  e ^ x\mape&{1\over\pi^3}
\int
d^4q{\Ga(1+N_0)\Ga(1+N_3)~[q]^{2J} \over
  (q_\ro P^2-m_0^2)^{1+N_0}(q_\ro P^2-m_3^2)^{1+N_3+J+c(J)}}
   e ^{xiq}\cr
&m_{0,3}\in\R,~~2J=0,1,\dots,~~N_{0,3}=0,1,\dots\end{array}\cr
 \end{eq}which
 arise
from the scalar irreducible ones $[m_0^2,m_3^2;0]$ by derivations 
with respect to the invariants 
${d\over dm_0^2}$, ${d\over dm_3^2}$
and - for the $\SO(3)/\SO(2)$-properties -
by derivations with respect to the 
spacetime variable
${d\over   d x}$.

\subsection
[Cartan Group Projection of Spacetime Representations]
{Cartan Group Projection\\of Spacetime Representations}

The projection  of the spacetime representations
to representations  of
Cartan subgroups  is given by
 {\it time $\D(\bl 1_2)$-projection via position space integration} and 
{\it boost $\SO^+(1,1)$-projection via time integration}
\begin{eq}{rlrl}
{\ep(x_0)\over 8i\pi}\int d^3 x :&\rep\D(2)&\map&\rep \D(\bl 1_2)\cr 
{1\over2}\int d x_0:&\rep\D(2)&\map&\rep \SD(2)\cr 
&&[{\rvec x\over |\rvec x|}]^{2J}
:&\rep\SD(2)\map\rep \SO^+(1,1) 
\end{eq}

The time projection for the irreducible representations
\begin{eq}{l}
{\ep(x_0)\over 8i\pi}\int d^3 x [m_0^2,m_3^2;2J](x)
={\ep(x_0)\over i\pi}\int
d^1q~{[q_0]^{2J} \over   (q_\ro P^2-m_0^2)(q_\ro P^2-m_3^2)^{1+J+c(J)}}
    e ^{x_0iq}\cr
\end{eq}can be computed with the $\SO(2)$-re\-pre\-sen\-ta\-tion matrix elements
\begin{eq}{l}
{\ep(x_0)\over i\pi}\int d^1 q{ {\scriptsize\pmatrix{ q \cr m\cr}}  \over
  q _{\ro P}^2-m^2}
 e^{x_0iq }
={1\over i\pi}\oint d^1 q{ {\scriptsize\pmatrix{ q \cr m\cr}}  \over
  q^2-m^2}
 e^{x_0iq }
={\scriptsize\pmatrix{
\cos x_0 m\cr
i\sin x_0 m\cr}}
\end{eq}The energy-momentum polynomials are projected to
energy polynomials
\begin{eq}{l}
\hbox{with }
\begin{array}{rl}
q_j&\mape \de_j^0 q_0\cr
g_{jk}&\mape \de_j^0\de_k^0\end{array}\then
\left\{\begin{array}{rl}
[q_0]^0&=1\cr
[q_0]^1&=q_0\cr
[q_0]^2&={3\over4}(q_0)^2,~~\dots\end{array}\right.
\end{eq}

The  projection to representations of the boost manifold
\begin{eq}{l}
{1\over2}\int d x_0
[m_0^2,m_3^2;2J](x)
={1\over \pi^2}\int
{d^3q~[q_a]^{2J}(-1)^{J+c(J)} \over   (\rvec q^2+m_0^2)
(\rvec q^2+m_3^2)^{1+J+c(J)}}
    e ^{-\rvec xi\rvec q}\cr
\end{eq}is computed with 
the Yukawa potential 
\begin{eq}{l}
{1\over \pi^2}\int d^3q {1\over \rvec q^2+m^2}e^{-\rvec xi\rvec q}
=2{e^{-|\rvec x m|}\over|\rvec x|},~~\rvec x\ne 0
\end{eq}which by itself is no 
$\SD(2)$-re\-pre\-sen\-ta\-tion.
The linear combinations occurring
in the $\SD(2)$-projection are measured $\SD(2)$-re\-pre\-sen\-ta\-tion
with finite spectral momenta for the measures, e.g.
\begin{eq}{l}
2{e^{-|\rvec x m_0|}-e^{-|\rvec x m_3|}\over|\rvec x|}
=\int _{m_0^2}^{m_3^2}dm^2{e^{-|\rvec x m|}\over |m|}
=\int_0^\infty dm^2\mu_0(m^2)[m^2;0](\rvec x)\cr
\mu_0(m^2)=\vth(m^2-m_0^2)\vth(m_3^2-m_0^2),~~
[m^2;0](\rvec x)={e^{-|\rvec x m|}\over |m|}\cr 
\int_0^\infty dm^2\mu_0(m^2)=m_3^2-m_0^2,\dots
\end{eq}The irreducible energy-momentum polynomials  are projected to
momentum polynomials $[\rvec q]^{2J}$, in general decomposable
\begin{eq}{l}
\hbox{with }
\begin{array}{rl}
q_j&\mape \de_j^a q_a\cr
g_{jk}&\mape -\de_j^a\de_k^b\de_{ab}\end{array}
\then\left\{
\begin{array}{rl}
[q_a]^0&=1\cr
[q_a]^1&=q_a\cr
[q_a]^2&=q_aq_b-{\de_{ab}\over4}\rvec q^2=
[\rvec q]^2+{\de_{ab}\over12}\rvec q^2[\rvec q]^0,~~
\dots\end{array}\right.
\end{eq}

\subsection{Scalar Spacetime Representations}

The {\it irreducible scalar spacetime representations} 
are 
\begin{eq}{l}
\D(2)\ni  e ^x\mape [m_0^2,m_3^2;0](x)
={1\over\pi^3}\int d^4q {1\over (q^2_{\ro P}-m_0^2)(q^2_{\ro P}-m_3^2)}
  e ^{xiq}\cr
\end{eq}

The decomposition
in energy-mo\-men\-ta measures with one singularity only
\begin{eq}{r}
{1\over(q^2- m_0^2)(q^2- m_3^2)}={1\over m_0^2-m_3^2}
[{1\over q^2- m_0^2}-{1\over q^2- m_3^2}]
\sim {\de(q^2- m_0^2)-\de(q^2- m_3^2)\over m_0^2-m_3^2}
\end{eq}gives the representation matrix elements for the time subgroup 
\begin{eq}{l}
\D(1)\ni   e ^{x_0}\mape {\ep(x_0)\over 8i\pi}\int d^3 x[m_0^2,m_3^2;0](x)
={i\over m_0^2-m_3^2}
[{\sin x_0 m_0\over m_0}-{\sin x_0 m_3\over m_3}]
\end{eq}The  boost group $\SO^+(1,1)$ is represented
with $J=0$ 
\begin{eq}{l}
\SO^+(1,1)\ni   e ^{-|\rvec x|}
\mape{1\over2} \int dx_0[m_0^2,m_3^2;0](x)
=-2{  e ^{-|\rvec xm_0|}-  e ^{-|\rvec xm_3|}\over|\rvec x|(m_0^2-m_3^2)}
\end{eq}

The explicit form of the irreducible scalar  spacetime representations
\begin{eq}{l}
[m_0^2,m_3^2;0](x)
=\vth(x^2)
{m_0^2\cl E_1({m_0^2x^2\over4})-m_3^2
\cl E_1({m_3^2 x^2\over4})\over  m_0^2-m_3^2}\cr
\end{eq}with  the special cases for  equal and trivial masses
\begin{eq}{rl}
[m^2,0;0](x)&=\vth(x^2)\cl E_1({m^2x^2\over4})\cr
[m^2,m^2;0](x)&=\vth(x^2)\cl E_0({m^2x^2\over4})\cr
[0,0;0](x)&=\vth(x^2)\cr
\end{eq}contain the measured $\D(1)$-re\-pre\-sen\-ta\-tions 
with Bessel functions $J_k$
\begin{eq}{l}
\cl E_k({\tau^2\over4})
={J_k(\tau)\over({\tau\over2})^k}
={\SUM_{n=0}^\infty}{ (-{\tau^2\over4})^n\over n!(n+k)!}
={1\over \sqrt\pi\Ga(k+{1\over2})}\int d E
\sqrt{1- E ^2}^{2k-1}\vth(1-E^2)  e ^{ \tau iE}\cr
k=0,1,\dots
\end{eq}

\subsection{Fundamental Spacetime Representations}

The irreducible {\it fundamental spacetime representations }  
belong to the 
generating real 4-di\-men\-sio\-nal $\SO^+(1,3)$-re\-pre\-sen\-ta\-tion
$[1|1]$
\begin{eq}{rl}
\D(2)\ni  e ^x\mape [m_0^2,m_3^2;1](x)
={1\over \pi^3}\int d^4q{q^j\si_j 
\over (q^2_{\ro P}-m_0^2)(q^2_{\ro P}-m_3^2)^2}
  e ^{xiq}\cr
\end{eq}They involve a simple pole (particle singularity)
and a dipole (interaction singularity)
reflecting the positive unitary  and the indefinite unitary representation of
a Cartan subgroup time $\D(1)$ and boost $\SO^+(1,1)$ resp.

The decomposition
into energy-mo\-men\-ta measures with one singularity only
\begin{eq}{rl}
{1\over(q^2- m_0^2)(q^2- m_3^2)^2}&=
{1\over (m_0^2-m_3^2)^2}
[{1\over q^2- m_0^2}-{1\over q^2- m_3^2}]
-{1\over (m_0^2-m_3^2)}
{1\over (q^2- m_3^2)^2}\cr
&\sim {\de(q^2- m_0^2)-\de(q^2- m_3^2)\over (m_0^2-m_3^2)^2}
+{\de' (q^2- m_3^2)\over m_0^2-m_3^2}\cr
\end{eq}gives the representation matrix elements for 
the time subgroup 
\begin{eq}{rl}
\D(1)\ni   e ^{x_0}\mape 
&-{\ep(x_0)\over 16\pi}\tr\int d^3 x [m_0^2,m_3^2;1](x)\cr
&={\cos x_0 m_0-\cos x_0 m_3\over (m_0^2-m_3^2)^2}
+{x_0m_3\sin x_0m_3\over 2m_3^2(m_0^2-m_3^2)}
\end{eq}and  those for the  
boost subgroup 
\begin{eq}{rl}
 \SO^+(1,1)\ni  e ^{-|\rvec x|}\mape&{1\over 4i}
 \tr{\rvec\si\rvec x\over|\rvec x|}\int dx_0
[m_0^2,m_3^2;1](x)\cr
&=2{(1+|\rvec x m_0|)  e ^{-|\rvec x m_0|}
-(1+|\rvec x m_3|)  e ^{-|\rvec x m_3|}
\over \rvec x^2( m_0^2- m_3^2)^2}
+{  e ^{-|\rvec x m_3| }
\over m_0^2- m_3^2}
\end{eq}

The integrated form of the spacetime representation 
\begin{eq}{l}
[m_0^2,m_3^2;1](x)=i\vth(x^2)x
{
 m_0^4\cl E_2({x^2 m_0^2\over4})
- m_3^4\cl E_2({x^2 m_3^2\over4})-( m_0^2- m_3^2) m_3^2
\cl E_1({x^2 m_3^2\over4})\over 2(m_0^2-m_3^2)^2}
\end{eq}has the special cases for  equal and trivial masses
\begin{eq}{rl}
[m^2,0;1](x)&=i\vth(x^2)x{\cl E_2({m^2x^2\over4})\over2}\cr
[0,m^2;1](x)&=i\vth(x^2)x
{\cl E_1({m^2x^2\over4})-\cl E_2({m^2x^2\over4})\over2}\cr
[m^2,m^2;1](x)&=i\vth(x^2)x{\cl E_0({m^2x^2\over4})\over4}\cr
[0,0;1](x)&=i\vth(x^2)x{1\over4}\cr
\end{eq}

\subsection{Spacetime  Quantum Fields}

Spacetime representations arise as field quantizations. In analogy to the
time dependent position $\bl x(t)$ 
quantized by a time $\D(1)$-re\-pre\-sen\-ta\-tion matrix
element, e.g. for the harmonic oscillator
\begin{eq}{l}
[m^2](t)={1\over \pi}
\int d^1 q{1\over q_\ro P^2-m^2}e^{tiq}
=-\ep(t){\sin tm\over m}=i\ep(t)[\bl x,\bl x](t)
\end{eq}the $\D(2)$-spacetime 
residual representations are quantizations of spacetime
fields, e.g. for the scalar and fundamental representation
as commutator and anticommutator of a scalar and spinor field resp.
\begin{eq}{rll}
[m_0^2,m_3^2;0](x)&={1\over\pi^3}
\int d^4q {1\over (q^2_{\ro P}-m_0^2)(q^2_{\ro P}-m_3^2)}  e ^{xiq}
&=i\ep(x_0)\com{\Phi}{\Phi}(x)\cr
[m_0^2,m_3^2;1](x)&={1\over\pi^3}
\int d^4q {q^j\si_j\over (q^2_{\ro P}-m_0^2)(q^2_{\ro P}-m_3^2)^2}  e ^{xiq}
&=i\ep(x_0)\acom{\Psi^*}{\Psi}(x)\cr
\end{eq}

In contrast to time and because of the
additional indefinite $\SO^+(1,1)$ boost structure such spacetime fields
cannot be interpreted in terms of positive metric particles only.
Supplementing the residual spacetime representation 
which can be taken  as a causally supported
quantization in flat tangent spacetime by an  Fock state 
value for the quantization opposite commutator, also spacelike supported 
\begin{eq}{rl}
\angle{\acom{\Phi}{\Phi}(x)
-\ep(x_0)[\Phi,\Phi](x)}
&={i\over \pi^3}
\int d^4q {1\over (q^2+io-m_0^2)(q^2_{\ro P}-m_3^2)}  e ^{xiq},~~m_0^2>m_3^2\cr
\angle{[\Psi^*,\Psi](x)-\ep(x_0)\acom{\Psi^*}{\Psi}(x)}
&={i\over \pi^3}
\int d^4q {q^j\si_j\over (q^2_{\ro P}+io -m_0^2)(q^2_{\ro P}-m_3^2)^2}  e ^{xiq}\cr
\end{eq}only the $m_0^2$-singularity with a positive residue
allows a particle interpretation and therefore an additional on-shell
 spacelike contribution as included with the integration prescription
$+io$.

Starting from a frequency $m$ for the 
creation operator $\ro u$ of an harmonic oscillator the 
freqencies $nm$ for the powers $\ro u^n$ with  natural $n$   
arise as singularities by convolutions 
of the basic representations $e^{tim}$ in the residual representation.
Similarily a fundamental spacetime representation $[m_0^2,m_3^2,1]$
may give rise to product representations whose positive metric singularities 
have a particle interpretation. To this end the convolution, appropriate for
the abelian time group $\D(1)=\GL(\C_\R)/\U(1)$, has to be 
generalized to a 
`convolution' for the nonabelian spacetime symmetric space  
$\D(2)=\GL(\C^2_\R)/\U(2)$.

As another genuine spacetime feature the class property
of the spacetime elements with the fixgroup $\U(2)$
 has to be taken into account 
\begin{eq}{l}
\GL(\C^2_\R)\cong \GL(\C^2_\R)/\U(2)\x\U(2)=\D(2)\x\U(2)
 \end{eq}This noncompact-compact factorization  can be connected with 
the external-internal dichotomy\cite{S982}.


\newpage
\begin{thebibliography}{99}

\bibitem{BOE}{H. Boerner, {\it Darstellungen von Gruppen} (1955), Springer,
      Berlin, G\" ot\-tin\-gen, Heidelberg}




\bibitem{ENS4}{N. Bourbaki, {\it Th\'eorie des Ensembles, Chapitre 4}
 (Structures) (1957),  Hermann,  Paris} 

\bibitem{LIE48}{N. Bourbaki, {\it Groupes et Alg\`ebres de Lie, Chapitres 4-8}
(1968-1975) , Hermann, Paris}



\bibitem{FOCK}{V. Fock, {\it Zeitschrift f\" ur Physik} 98 (1935), 145}

\bibitem{FINK}{D.R. Finkelstein, {\it Quantum Relativity} (1996), Springer}


\bibitem{FULHAR}{W. Fulton, J. Harris, {\it Representation Theory} (1991), Springer}
 

\bibitem{GELVIL}{I.M. Gel'fand, M.I. Graev, N.Ya. Vilenkin,
{\it Generalized Functions V (Integral Geometry and Representation Theory)}
(1962, English translation 1966), Academic Press, New York and London}

\bibitem{GELNAI}{I.M. Gelfand, M.A. Neumark,
{\it Unit\" are Darstellungen der klassischen Gruppen}
(1950, German Translation 1957), Akademie Verlag, Berlin}


 
\bibitem{HEL}{S. Helgason, {\it Differential Geometry, Lie Groups and
       Symmetric Spaces} (1978), Academic Press, 
        New York. London, Toronto, Sidney, San Francisco}
 


\bibitem{KL}{G. K\" allen, {\it Helvetica Physica Acta}, 25 (1952), 417;
H. Lehmann, {\it Nuovo Cimento} 11 (1954), 342}




\bibitem{WEYLRZM}{H. Weyl, {\it Raum-Zeit-Materie} (1923),
Wis\-sen\-schaft\-li\-che Buch\-ge\-sell\-schaft, Darmstadt}


\bibitem{WIG}{E. P. Wigner, {\it Annals of Mathematics} 40 (1939), 149}

\bibitem{S89}{H. Saller, {\it Nuovo Cimento} 104B (1989), 291}


\bibitem{S922}{H. Saller, {\it Nuovo Cimento} 108B (1993), 603,
and 109B (1993), 255}
 


\bibitem{SBH95}{H. Saller,  R. Breuninger and M. Haft,
{\it Nuovo Cimento} 108A (1995), 1225}



\bibitem{S97}{H. Saller,
{\it International Journal of Theoretical Physics} 36 (1997), 2783}


\bibitem{S981}{H. Saller,
The Central Correlations of Hypercharge, 
Isospin, Colour and Chirality in the
Standard Model, hep-th/9802112, to be published}

\bibitem{S982}{H. Saller,
{\it International Journal of Theoretical Physics} 37 (1998), 2333}



 \end{thebibliography}
\end{document}